\documentstyle[11pt]{article}
\normalbaselines

\newtheorem{lem}{Lemma}[section]
\newtheorem{prop}{Proposition}[section]

\begin{document}

\def\bea*{\begin{eqnarray*}}
\def\eea*{\end{eqnarray*}}
\def\ba{\begin{array}}
\def\ea{\end{array}}
\count1=1
\def\be{\ifnum \count1=0 $$ \else \begin{equation}\fi}
\def\ee{\ifnum\count1=0 $$ \else \end{equation}\fi}
\def\ele(#1){\ifnum\count1=0 \eqno({\bf #1}) $$ \else \label{#1}\end{equation}\fi}
\def\req(#1){\ifnum\count1=0 {\bf #1}\else \ref{#1}\fi}
\def\bea(#1){\ifnum \count1=0   $$ \begin{array}{#1}
\else \begin{equation} \begin{array}{#1} \fi}
\def\eea{\ifnum \count1=0 \end{array} $$
\else  \end{array}\end{equation}\fi}
\def\elea(#1){\ifnum \count1=0 \end{array}\label{#1}\eqno({\bf #1}) $$
\else\end{array}\label{#1}\end{equation}\fi}
\def\cit(#1){
\ifnum\count1=0 {\bf #1} \cite{#1} \else 
\cite{#1}\fi}
\def\bibit(#1){\ifnum\count1=0 \bibitem{#1} [#1    ] \else \bibitem{#1}\fi}
\def\ds{\displaystyle}
\def\hb{\hfill\break}
\def\comment#1{\hb {***** {\em #1} *****}\hb }

\newcommand{\TZ}{\hbox{\bf T}}
\newcommand{\MZ}{\hbox{\bf M}}
\newcommand{\ZZ}{\hbox{\bf Z}}
\newcommand{\NZ}{\hbox{\bf N}}
\newcommand{\RZ}{\hbox{\bf R}}
\newcommand{\CZ}{\,\hbox{\bf C}}
\newcommand{\PZ}{\hbox{\bf P}}
\newcommand{\QZ}{\hbox{\rm eight}}
\newcommand{\HZ}{\hbox{\bf H}}
\newcommand{\EZ}{\hbox{\bf E}}
\newcommand{\GZ}{\,\hbox{\bf G}}

\font\germ=eufm10
\def\goth#1{\hbox{\germ #1}}
\vbox{\vspace{38mm}}

\begin{center}
{\LARGE \bf Quantum Group Theory in $\tau^{(2)}$-model, Duality of $\tau^{(2)}$-model and XXZ-model with Cyclic ${\bf U_q(sl_2)}$-representation for ${\bf q^n =1}$, and Chiral Potts Model} \\[10 mm] 
Shi-shyr Roan \\
{\it Institute of Mathematics \\
Academia Sinica \\  Taipei , Taiwan \\
(email: maroan@gate.sinica.edu.tw ) } \\[25mm]
\end{center}
We identify the quantum group ${\Large\textsl{U}}_\textsl{w}(sl_2)$ in the $L$-operator of $\tau^{(2)}$-model for a generic $\textsl{w}$ as a subalgebra of $U_{\sf q} (sl_2)$ with $\textsl{w} = {\sf q}^{-2}$. In the roots of unity case, ${\sf q}=q, \textsl{w} = \omega$ with $q^{{\bf n}} = \omega^N = 1$,  the eigenvalues and eigenvectors of XXZ-model with the $U_q (sl_2)$-cyclic representation are determined by the $\tau^{(2)}$-model with the induced ${\Large\textsl{U}}_\omega(sl_2)$-cyclic representation, which is decomposed as a finite sum of $\tau^{(2)}$-models in non-superintegrable inhomogeneous $N$-state chiral Potts model. Through the theory of chiral Potts model, the $Q$-operator of XXZ-model can be identified with the related chiral Potts transfer matrices, with special features appeared in the ${\bf n}=2N$, e.g. $N$ even, case.
We also establish the duality of $\tau^{(2)}$-models related to cyclic representations of $U_q (sl_2)$, analogous to the $\tau^{(2)}$-duality in chiral Potts model; and identify the model dual to the XXZ model with $U_q (sl_2)$-cyclic representation. 
\begin{abstract}

\end{abstract}
\par \vspace{5mm} \noindent
{\rm 2008 PACS}:  05.50.+q, 02.20.Uw, 75.10Pq \par \noindent

\setcounter{section}{0}
\section{Introduction}
\setcounter{equation}{0}
The $\tau^{(2)}$-model is the six-vertex model first appeared in the $N$-state chiral Potts model (CPM) in \cite{BazS} as an adjacent model for the study of chiral Potts transfer matrix in the frame work of the Baxter's $TQ$-relation \cite{Bax}. Together with a set of $\tau^{(j)}$-models and functional relations among the various transfer matrices \cite{BPA}, one can determine all the eigenvalues of both matrices, hence solve the eigenvalue problem of CPM \cite{AMP, B90, B91, BBP, MR, R0805}. One also expects the eigenvector problem of CPM should equally rely on the $\tau^{(2)}$-eigenvectors as well, as already revealed in the relation of state-correspondence in the duality of chiral Potts model \cite{R09}. Furthermore, the recent progress made on the eigenvector problem in superintegrable $N$-state CPM for odd $N$ in \cite{R10} again showed the vital role of $\tau^{(2)}$-model about the degeneracy symmetries of $\tau^{(2)}$-eigenspaces through its equivalent spin-$\frac{N-1}{2}$ XXZ-chain. 
It is known that the theory of quantum group $U_{\sf q} (sl_2)$ for a generic ${\sf q}$ stemmed from the study of  XXZ-model as an equivalent formulation of the Yang-Baxter (YB) relation (\req(YBXXZ)) of $L$-operator \cite{Fad, KRS}.  
In the case when ${\sf q}$ is an $N$th root of unity for odd $N$, the XXZ chain with cyclic representation was  identified with the chiral Potts $\tau^{(2)}$-model in \cite{R0806, R09}. The observation and arguments there can be further extended  to a general setting for a quantum group $U_{\sf q} (sl_2)$ with a generic ${\sf q}$. In this work, we find a quantum subalgebra ${\Large\textsl{U}}_\textsl{w}(sl_2)$ of $U_{\sf q} (sl_2)$ for a generic $\textsl{w} ~ (= {\sf q}^{-2})$ (see (\req(qUw)) in the paper), with an associated $L$-operator satisfying the YB relation (\req(YBt2)) of $\tau^{(2)}$-model. The understanding of local state vectors of a statistical $\tau^{(2)}$-model  is thus reduced to the representation theory of ${\Large\textsl{U}}_\textsl{w}(sl_2)$.  
The theory of $N$-state CPM is in essence the study of $Q$-operator associated to XXZ-chains with the cyclic representation in the root of unity case: ${\sf q}=q, ~ \textsl{w} = \omega$ a primitive ${\bf n}$th and $N$th root of unity for ${\bf n}  ~(\geq 3), N  ~( \geq 2)$ respectively with the relation
\bea(lll)
q^{-2} = \omega, & q^{{\bf n}} = \omega^N = 1, & {\bf n}= N ~ ~ {\rm odd ,~ ~ or ~ ~ } {\bf n} = 2N.
\elea(qomega)
The aim of this paper is to find an explicit relationship between XXZ-chains of $U_q (sl_2)$-cyclic representation and non-superintegrable $N$-state chiral Potts model, with especial attention on even $N$ case; and explore the duality theory connected to XXZ-models with cyclic representation. Since every cyclic representation of $U_q (sl_2)$ induces a representation of ${\Large\textsl{U}}_\omega (sl_2)$, the XXZ-model with an $U_q (sl_2)$-cyclic representation gives rise to a $\tau^{(2)}$-model, denoted by $\textsc{t}^{(2)}(t)$ in (\req(XXZcyc)), with the induced  ${\Large\textsl{U}}_\omega (sl_2)$-cyclic representation. Correspondingly,  the ABCD-algebra, i.e. the monodromy algebra, of the XXZ-model is generated by ABCD-algebra of $\textsc{t}^{(2)}$-model and $K^\frac{1}{2}$. Hence one can study the XXZ-model with a cyclic representation through its induced $\textsc{t}^{(2)}$-model. By the representation theory of ${\Large\textsl{U}}_\omega (sl_2)$, we find that a $\textsc{t}^{(2)}$-model with the chain-size $L$ is decomposed as $({\bf n}/N)^L$ sub-models, each of which is isomorphic to some chiral-Potts $\tau^{(2)}$-model with inhomogeneous vertical rapidities. The $K^\frac{1}{2}$-operator of XXZ-model gives rise to a pairing of these sub-models with identical eigenvalues and eigenvectors, also with similar $Q$-operators. By this, the eigenvalues and eigenvectors of XXZ-model are obtained , and the $Q$-operator can be identified with the chiral Potts transfer matrix.  
In the duality theory of CPM in \cite{R09}, $\tau^{(2)}$-duality is the equality of two  $\tau^{(2)}$-models by  identifying the "ordered- and disordered-" state vectors through the spin-and-face-variable-expression of a $\tau^{(2)}$-transfer matrix \cite{B93}. We would like to examine how far the $\tau^{(2)}$-duality in CPM can be extended to XXZ-models with cyclic representation. First, we notice that the $\tau^{(2)}$-duality of CPM for the sub-models in the decomposition of $\textsc{t}^{(2)}(t)$ can not be carried over to XXZ model, partly due to the fact that the duality works on transfer matrices only, not as ABCD-algebra- or ${\Large\textsl{U}}_\omega (sl_2)$-representations. However, we are still able to find a spin-and-face-expression of the transfer matrix of XXZ model with cyclic representation, and identify the model dual to it. It turns out the dual model of XXZ-chains with cyclic representation is an another type of $\tau^{(2)}$-model, $\textsc{t}^{\dagger (2)}(t)$ in (\req(t2dag)), dual to the $\textsc{t}^{(2)}(t)$ inherited from  XXZ-model. This shows the dualities involved with XXZ-models all depend on the duality of $\tau^{(2)}$-models, which we  expect also serves the fundamental role in the duality of XXZ-model with a representation of $U_q (sl_2)$ other than cyclic representations.

This paper is organized as follows. In section \ref{sec.t2qunt}, we provide a ${\Large\textsl{U}}_\textsl{w}(sl_2)$-quantum group formulation of $\tau^{(2)}$-model for a generic ${\sf w}$, as a parallel theory to the quantum group $U_{\sf q}(sl_2)$ in XXZ-chains for a generic ${\sf q}$. We then derive some basic properties about the monodromy-(ABCD)- algebras of the general $\tau^{(2)}$-model. Afterwards, we shall concern only with the root of unity case (\req(qomega)) for the rest of this paper.  Section \ref{sec.CyclMod} is devoted to the study of $\tau^{(2)}$-model and XXZ-model with cyclic representation (of ${\Large\textsl{U}}_\omega (sl_2)$ and $U_q ( sl_2)$ respectively). In subsection \ref{ssec.L}, we first recall the cyclic $\CZ^{\bf n}$-representation of $U_q (sl_2)$, and describe its related cyclic representations of  ${\Large\textsl{U}}_\omega (sl_2)$; then present a detailed structure about the representation theory of $L$-operators in $\tau^{(2)}$-model and XXZ-model with cyclic representation. In subsection \ref{ssec.XXZc}, we study the structure of XXZ-model and $\tau^{(2)}$-model induced from  cyclic $\CZ^{\bf n}$-representation of $U_q (sl_2)$. Each irreducible component of $\tau^{(2)}$-model with the induced cyclic $\CZ^{\bf n}$-representation is isomorphic to some $\tau^{(2)}$-model associated to $N$-state CPM with (possible) inhomogeneous vertical rapidities. By this, we are able to find the connection about eigenvalues and eigenvectors between XXZ-model and $\tau^{(2)}$-model in CPM. 
Section \ref{sec.t2dual} is devoted to the discussion of  duality of $\tau^{(2)}$-models and XXZ models with cyclic representation. First in subsection \ref{ssec.tau2dual}, we recall the $\tau^{(2)}$-duality in CPM; and then in subsection \ref{ssec.t2dual}, we study the duality of $\tau^{(2)}$-model with cyclic $\CZ^{\bf n}$-representation, and find the duality between $\textsc{t}^{(2)}$- and $\textsc{t}^{\dagger (2)}$-models.
In subsection \ref{ssec.Relt2}, we examine the relationship between the $\tau^{(2)}$-dualities found in the previous two subsections. The $\textsc{t}^{(2)}$-$\textsc{t}^{\dagger (2)}$-duality agrees with the $\tau^{(2)}$-duality in CPM only in the ${\bf n}= N$ odd case, but differs in ${\bf n}= 2N$ case. In subsection \ref{ssec.XXZdual}, we identify the $\textsc{t}^{\dagger (2)}$-model as the dual model of XXZ model with cyclic representation using a spin-and-face-expression of the XXZ-transfer matrix. 
Section \ref{sec.CPM} is the discussion of the relationship between CPM and XXZ model with $U_q (sl_2)$-cyclic representation. First, we recall some basic notions and the duality in CPM in subsection \ref{ssec.duCPM}. Then in subsection \ref{ssec.CPXXZ}, we identify the chiral Potts transfer matrix as the $Q$-operator of XXZ model with cyclic representation.

{\bf Notation}:  In this paper, we use the following standard notations. For a positive integer $N$ greater than one, 
$\CZ^N$ denotes the vector space of $N$-cyclic vectors with the canonical base 
$|\sigma \rangle, \sigma  \in \ZZ_N ~ (:= \ZZ/N\ZZ)$. For a $N$th primitive root of unity $\omega$, e.g. $\omega = {\rm e}^{\frac{2 \pi {\rm i}}{N}}$, the Weyl operators $X, Z$ (with respective to $\omega$) with the relations, $X^N=Z^N=1$ and $XZ= \omega^{-1}ZX$, are defined by
$$
 X |\sigma  \rangle = | \sigma  +1 \rangle , ~ \ ~ Z |\sigma  \rangle = \omega^\sigma  |\sigma  \rangle ~ ~ \ ~ ~ (\sigma  \in \ZZ_N) .
$$
The Fourier basis $\{ \widehat{|k } \rangle \}$ of $\{ | \sigma  \rangle \}$ is defined by 
\be
\widehat{|k } \rangle  = \frac{1}{\sqrt{N}} \sum_{\sigma =0}^{N-1} \omega^{-k \sigma} |\sigma \rangle , ~ ~ ~ | \sigma  \rangle = \frac{1}{\sqrt{N}} \sum_{k =0}^{N-1} \omega^{\sigma k} \widehat{|k} \rangle , ~ \sigma  \in \ZZ_N,
\ele(Fb) 
with the corresponding Weyl operators, $
\widehat{X} \widehat{|k} \rangle = \widehat{|k+1} \rangle$, $\widehat{Z} \widehat{|k} \rangle = \omega^k \widehat{|k} \rangle$ satisfying $\widehat{X}\widehat{Z}= \omega^{-1}\widehat{Z}\widehat{X}$. Then the following equality holds:
\be
(X, Z) = (\widehat{Z}, \widehat{X}^{-1}). 
\ele(XZF)
The Fourier bases of $\stackrel{L}{\otimes} \CZ^N$ are denoted by 
$$
\begin{array}{llll}
|\sigma_1, \ldots, \sigma_L  \rangle := | \sigma_1  \rangle \otimes \ldots, \otimes \sigma_L , & {\rm or} & |\widehat{k}_1, \ldots, \widehat{k}_L  \rangle := \widehat{|k_1 } \rangle \otimes \ldots, \otimes \widehat{|k_L } \rangle , &  (\sigma_j, k_j \in \ZZ_N).
\end{array} 
$$

\section{ Quantum Group Theory of  $\tau^{(2)}$-model \label{sec.t2qunt}}
\setcounter{equation}{0}
For a generic ${\sf q} \in \CZ \setminus \{0, \pm 1 \}$, we denote $\textsl{w} = {\sf q}^{-2}$. It is well-known that
the quantum group $U_{\sf q} (sl_2)$ is the algebra generated by $K^\frac{\pm 1}{2}, e^\pm$ with the relations $K^\frac{\pm 1}{2}K^\frac{\mp 1}{2}=1 $ and 
\be 
 K^{\frac{1}{2}} e^{\pm } K^\frac{-1}{2}  =   {\sf q}^{\pm 1} e^{\pm}  , ~ 
~ [e^+ , e^- ] = \frac{K-K^{-1}}{{\sf q} - {\sf q}^{-1}}. 
\ele(Uq)
Define the $L$-operator with non-zero complex parameter $\rho , \nu \in \CZ$:
\bea(lll)
{\bf L} (s)  &(= {\bf L} (s ; \rho , \nu )) & =  \left( \begin{array}{cc}
         \rho^{-1} \nu^\frac{1}{2} s K^\frac{-1}{2}   -  \nu^\frac{-1}{2} s^{-1} K^\frac{1}{2}   &  ({\sf q} - {\sf q} ^{-1}) e^-    \\
        ({\sf q}  - {\sf q} ^{-1}) e^+ &  \nu^\frac{1}{2} s K^\frac{1}{2} -  \rho \nu^\frac{-1}{2} s^{-1} K^\frac{-1}{2} 
\end{array} \right).  
\elea(6vL)
The quantum group $U_{\sf q} (sl_2)$ is characterized by 
the YB relation,
\be
{\bf R}  (s/s') ({\bf L}(s) \bigotimes_{aux}1) ( 1
\bigotimes_{aux} {\bf L}(s')) = (1
\bigotimes_{aux} {\bf L}(s'))( {\bf L}(s)
\bigotimes_{aux} 1) {\bf R} (s/s'),
\ele(YBXXZ)
with the symmetric $R$-matrix \cite{Fad, KRS}:
$$
{\bf R} (s) = \left( \begin{array}{cccc}
        s^{-1} {\sf q}  - s {\sf q} ^{-1}  & 0 & 0 & 0 \\
        0 &s^{-1} - s & {\sf q}  - {\sf q} ^{-1} &  0 \\ 
        0 & {\sf q}  -{\sf q} ^{-1} &s^{-1} - s & 0 \\
     0 & 0 &0 & s^{-1} {\sf q}  - s {\sf q} ^{-1} 
\end{array} \right) .    
$$   
By setting $t = s^2$,  the modified $L$-operator, $-s \nu^\frac{1}{2} K^{\frac{-1}{2}} {\bf L} (s)$ with the gauge transform ${\rm dia}[1, -s\nu^\frac{1}{2}q]$, is expressed by 
\bea(lll)
 {\Large\textsl{L}} (t) &(= {\Large\textsl{L}} (t; ; \rho , \nu) ) & = \left( \begin{array}{cc}
        1- t \nu  \rho^{-1} K^{-1}  &  (1- \textsl{w} ) E^-    \\
         - t \nu (1- \textsl{w}) E^+ & - t \nu   +    \rho K^{-1} 
\end{array} \right) ,   
\elea(Ltau)
where $E^+ = - {\sf q}^2 K^\frac{-1}{2} e^+, E^- = K^\frac{-1}{2} e^-$.  The quantum subalgebra of $U_{\sf q} (sl_2)$ generated by  $K^{\pm 1}, E^\pm$, with the generator-relation: 
\bea(ll)
KE^\pm K^{-1} = \textsl{w}^{\mp 1}  E^\pm , & \textsl{w} E^+ E^- - E^- E^+ = \frac{K^{-2}-1 }{1-\textsl{w}},
\elea(qUw)
will be denoted by ${\Large\textsl{U}}_\textsl{w}(sl_2)$. 
Then the $L$-matrix in (\req(Ltau)) satisfy the YB equation,
\be
\textsl{R}(t/t') ({\Large\textsl{L}} (t) \bigotimes_{aux}1) ( 1
\bigotimes_{aux} {\Large\textsl{L}} (t')) = (1
\bigotimes_{aux} {\Large\textsl{L}} (t'))({\Large\textsl{L}} (t)
\bigotimes_{aux} 1) \textsl{R}(t/t') 
\ele(YBt2)
with the asymmetry $R$-matrix
$$
\textsl{R}(t) = \left( \begin{array}{cccc}
        t \textsl{w} - 1  & 0 & 0 & 0 \\
        0 &t-1 & \textsl{w} - 1 &  0 \\ 
        0 & t(\textsl{w}  - 1) &( t-1)\textsl{w} & 0 \\
     0 & 0 &0 & t \textsl{w} - 1    
\end{array} \right).
$$
Indeed, the generator-relation (\req(qUw)) of ${\Large\textsl{U}}_\textsl{w}(sl_2)$ is characterized by the YB relation (\req(YBt2)) of the $L$-operator (\req(Ltau)). Hence a representation of $U_{\sf q} (sl_2)$ or ${\Large\textsl{U}}_\textsl{w}(sl_2)$ is equivalent to a representation of the $L$-operator (\req(6vL)) or (\req(Ltau)) satisfying the respective YB relation. Two representations of $L$-operator (\req(6vL)) or (\req(Ltau)) are equivalent if and only if the  induced equivalent representations of $U_{\sf q} (sl_2)$ or ${\Large\textsl{U}}_\textsl{w}(sl_2)$ are equivalent with the same parameter $\rho, \nu$. In this paper, we shall also make the identification of the quantum-algebra representation and $L$-operator representation when no confusion could arise. 

For a chain of size $L$, we assign the $L$-operator (\req(6vL)) or (\req(Ltau)) at the $\ell$th site, and form the XXZ-monodromy matrix 
\bea(ll)
\bigotimes_{\ell=1}^L {\bf L}_\ell (s)  =  \left( \begin{array}{cc}
        {\bf A} (s)  & {\bf B} (s) \\
        {\bf C} (s) & {\bf D} (s)
\end{array} \right) , & {\bf L}_\ell (s)= {\bf L} (s ; \rho_\ell , \nu) ~ {\rm at ~ site ~ } \ell ;
\elea(MXXZ)
and $\tau^{(2)}$-monodromy matrix 
\bea(ll)
\bigotimes_{\ell=1}^L {\Large\textsl{L}}_\ell (t)  =  \left( \begin{array}{cc}
        {\Large\textsl{A}} (t)  & {\Large\textsl{B}} (t) \\
        {\Large\textsl{C}} (t) & {\Large\textsl{D}} (t)
\end{array} \right) , & {\Large\textsl{L}}_\ell (t)= {\Large\textsl{L}} (t ; \rho_\ell , \nu ) ~ {\rm at ~ site ~ } \ell.
\elea(Mtau)
By the relation between $L$-operators (\req(6vL)) and (\req(Ltau)), the monodromy entries of (\req(MXXZ)) and (\req(Mtau)) are related by
\bea(ll)
{}^{{\Large\textsl{A}} (t)}_{{\Large\textsl{D}}(t)} = (-s)^L \nu^\frac{L}{2} K^\frac{-1}{2} {}^{{\bf A} (s)}_{{\bf D} (s)}, & {}^{{\Large\textsl{B}} (t)}_{ {\Large\textsl{C}} (t)} = (-s)^{L\mp 1} \nu^\frac{L \mp 1}{2} {\sf q}^{\mp 1} K^\frac{-1}{2} {}^{{\bf B} (s)}_{{\bf C} (s)},
\elea(tTMon)
where $K^\frac{1}{2}:=\stackrel{L}{\bigotimes} K_\ell^\frac{1}{2}$. Equivalently to say, the XXZ-(ABCD)-algebra (\req(MXXZ)) is generated by the $\tau^{(2)}$-(ABCD)-algebra (\req(Mtau)) and $K^\frac{1}{2}$, with the relation (\req(tTMon)) between  their algebra-generators. For an integer $r \in \ZZ$, the YB relation of the  monodromy matrices yields the commutating relation of the $r$-twisted traces:
\bea(ll)
{\bf T} (s) = {\bf A} (s) + {\sf q}^{-2r} {\bf D} (s), & \tau^{(2)} (t) = {\Large\textsl{A}} (\textsl{w} t)  + \textsl{w}^r {\Large\textsl{D}} (\textsl{w} t) 
\elea(Tranf)
for $s \in \CZ$ and $t \in \CZ$, as a one-parameter family in $\stackrel{L}{\bigotimes} U_{\sf q} (sl_2)$ or $\stackrel{L}{\bigotimes} {\Large\textsl{U}}_\textsl{w} (sl_2)$ respectively. Then we have the relation, $
[{\bf T} (s) , K^\frac{1}{2}] = [\tau^{(2)} (t), K]=0$, which indeed follow from the commutative relations of $K^\frac{1}{2}, K$ and monodromy entries in (\req(MXXZ)) or (\req(Mtau)):
$$ 
\begin{array}{llll} 
 K^{\frac{1}{2}} {}^{{\bf A} (s)}_{{\bf D} (s)} K^\frac{-1}{2} = {}^{{\bf A} (s)}_{{\bf D} (s)},  & K^{\frac{1}{2}} {}^{{\bf C} (s)}_{{\bf B} (s)} K^\frac{-1}{2}  =   {}^{{\sf q} {\bf C} (s)}_{{\sf q}^{-1}{\bf B} (s)}; &
K {}^{{\Large\textsl{A}} (t)}_{ {\Large\textsl{D}} (t)} K^{-1} = {}^{{\Large\textsl{A}} (t)}_{ {\Large\textsl{D}} (t)},  & 
K {}^{{\Large\textsl{C}} (t)}_{ {\Large\textsl{B}} (t)} K^{-1} = {}^{\textsl{w}^{-1} {\Large\textsl{C}} (t)}_{ \textsl{w} {\Large\textsl{B}} (t)}.
\end{array}
$$
By (\req(tTMon)), the $r$-traces of $\tau^{(2)}$ and ${\bf T}$ in (\req(Tranf)) are related by 
\bea(ll)
\tau^{(2)} ( t) = (-{\sf q}^{-1}s)^L \nu^\frac{L}{2} K^\frac{-1}{2} {\bf T} ({\sf q}^{-1} s), & t = s^2 .
\elea(tauT)
For given $\rho_\ell , \nu$, a $U_{\sf q} (sl_2)$-representation on $\CZ^d$ gives rise to a commuting family of  transfer matrices ${\bf T} (s)$ or $\tau^{(2)}(t)$ on $\stackrel{L}{\bigotimes} \CZ^d$ with the $L$-operator induced from (\req(6vL)) or (\req(Ltau)) respectively. In particular, when $\rho_\ell = 1,  \nu = {\sf q}^{d-2}$,  the spin-$\frac{d-1}{2}$ (highest weight) representation of $U_{\sf q} (sl_2)$ on  $\CZ^d = \oplus_{k=0}^{d-1} \CZ {\bf e}^k $:
\bea(lll)
K^{\frac{1}{2}} ({\bf e}^k) = {\sf q}^{\frac{d-1-2k}{2}} {\bf e}^k , & e^+ ( {\bf e}^k ) = [ k  ]_{\sf q} {\bf e}^{k-1} , & e^-( {\bf e}^k ) = [ d-1-k ]_{\sf q} {\bf e}^{k+1}, \\

K ({\bf e}^k) = \textsl{w}^{\frac{-d+1}{2}+ k} {\bf e}^k , & E^+ ( {\bf e}^k ) =  - \textsl{w}^{\frac{d-1}{2}-k} [k] {\bf e}^{k-1} ,& E^- ( {\bf e}^k )= \textsl{w}^\frac{-d+1}{2} [d-1-k]  {\bf e}^{k+1} , 
\elea(spinrp)
where $[n]_{\sf q}:= \frac{{\sf q}^n - {\sf q}^{-n}}{{\sf q}- {\sf q}^{-1}}, [n]:= \frac{1-\textsl{w}^n}{1-\textsl{w}}$, for the XXZ chain (\req(MXXZ)) gives rise to the well-known homogeneous XXZ model of spin-$\frac{d-1}{2}$ (see, e.g. \cite{KiR, R06Q, R06F} and references therein).  Note that in (\req(spinrp)), $ e^+ ( {\bf e}^{0} ) = e^- ( {\bf e}^{d-1} )= E^+ ( {\bf e}^{0} ) = E^- ( {\bf e}^{d-1} )= 0$. \par \vspace{.1in} \noindent
{\bf Remark.} In (\req(MXXZ)) and (\req(Mtau)), the $L$-operator, ${\bf L}_\ell (s)$ and ${\Large\textsl{L}}_\ell (t)$, are assumed with the same value of the parameter $\nu$ in (\req(6vL)) and (\req(Ltau)). However, one may also consider monodromy matrix defined by ${\bf L}_\ell (s)= {\bf L} (s ; \rho_\ell , \nu_\ell)$ and ${\Large\textsl{L}}_\ell (t)= {\Large\textsl{L}} (t ; \rho_\ell , \nu_\ell)$ with $\rho_\ell, \nu_\ell$'s distinct, and form the transfer matrix ${\bf T} (s ; \{\rho_\ell, \nu_\ell\}), \tau^{(2)} (t ; \{\rho_\ell, \nu_\ell\})$ as in (\req(Tranf)).
Since the $L$-operators in (\req(6vL)) and (\req(Ltau)) satisfy the relations,
$$
\begin{array}{lll}
{\bf L} (s ; \rho_\ell , \nu_\ell) = {\bf L} (\xi_\ell^\frac{1}{2} s ; \rho_\ell , \nu ),  &  {\Large\textsl{L}} ( t ; \rho_\ell , \nu_\ell ) =  {\Large\textsl{L}} (\xi_\ell t ; \rho_\ell , \nu ),  &  \nu_\ell = \xi_\ell \nu ,
\end{array}
$$
the transfer matrices ${\bf T} ( s ; \{\rho_\ell, \nu_\ell\}), \tau^{(2)} (t ; \{\rho_\ell, \nu_\ell\})$ are reduced to those in (\req(Tranf)) with the same $\nu$ by
\bea(ll)
{\bf T} (s ; \{\rho_\ell, \nu_\ell\}) = {\bf T} (\xi_1^\frac{1}{2}s, ... , \xi_L^\frac{1}{2}s), &  {\bf T} (s)= {\bf T} (s, ... ,s); \\
 \tau^{(2)} (t ; \{\rho_\ell, \nu_\ell\})= \tau^{(2)} (\xi_1 t, ... , \xi_L t), & \tau^{(2)} (t)= \tau^{(2)} (t, ... ,t).
\elea(Ttnul)

\section{XXZ-model and $\tau^{(2)}$-model with Cyclic Representation \label{sec.CyclMod}}
\setcounter{equation}{0}
In the root of unity case (\req(qomega)), we let $\widehat{|\sigma} \rangle$'s or $\widehat{|k} \rangle$'s  $(\sigma, k \in \ZZ_N )$ of $\CZ^N$ be the Fourier basis in (\req(Fb)), and  $(X, Z), (\widehat{X}, \widehat{Z})$ the Weyl operators (with respective to $\omega$), in (\req(XZF)). Similarly, for the cyclic ${\bf n}$-space $\CZ^{\bf n}$, the Fourier bases (with respective to $q$) will be denoted by 
\be
\widehat{|k } \rangle \rangle  = \frac{1}{\sqrt{\bf n}} \sum_{\sigma =0}^{{\bf n}-1} q^{-k \sigma} |\sigma \rangle \rangle , ~ ~ ~ | \sigma  \rangle \rangle = \frac{1}{\sqrt{\bf n}} \sum_{k =0}^{{\bf n}-1} q^{\sigma k} \widehat{|k} \rangle \rangle , ~ k, \sigma  \in \ZZ_{\bf n},
\ele(Fbq)
with the Weyl operators $(X', Z'), (\widehat{X}', \widehat{Z}')$ defined by
$$
\begin{array}{ll}
 X' | \sigma  \rangle \rangle = | \sigma + 1 \rangle \rangle , Z' | \sigma  \rangle \rangle = q^k | \sigma  \rangle \rangle ; & \widehat{X}' \widehat{|k} \rangle \rangle = \widehat{|k+1} \rangle \rangle , \widehat{Z}' \widehat{|k} \rangle \rangle = q^k \widehat{|k} \rangle \rangle .
\end{array}
$$
Then the relations  $X'Z' = q^{-1} Z' X'$ , $\widehat{X}'\widehat{Z}' = q^{-1} \widehat{Z}' \widehat{X}'$, and the identity
$(X', Z') = (\widehat{Z}', \widehat{X}'^{-1})$ hold. 
First, we describe some special cyclic $\CZ^N$-subspaces of $\CZ^{\bf n}$ for later use. 
Denote
$$ 
c({\bf n}) = 2^\frac{-1-(-1)^{\bf n}}{4} = \left\{\begin{array}{ll} 1 & {\rm if} ~ {\bf n}=N ~ {\rm odd} \\ 2^{-1/2} & {\rm if} ~ {\bf n}=2N. \end{array} \right.
$$

\begin{lem}\label{lem:CnN} Consider the cyclic $N$-vectors in $\CZ^{\bf n}$,
$$
 |\sigma \rangle \rangle_+ := c({\bf n}) |-2 \sigma \rangle \rangle, ~ |\sigma \rangle \rangle_- := c({\bf n}) |-2 \sigma +1 \rangle \rangle \in \CZ^{\bf n} ~  ~ ~ (\sigma \in \ZZ_N),
$$
and the linear transformations
\bea(lll)
\wp_\pm : \CZ^{\bf n} \longrightarrow \CZ^N , ~ & \wp_+( \widehat{|k} \rangle \rangle )= \widehat{|k} \rangle , &
\wp_-( \widehat{|k} \rangle \rangle )= q^{-k} \widehat{|k} \rangle ~ ~ ( k \in \ZZ_{\bf n}).
\elea(wp)
Define the cyclic $N$-subspaces ${\cal C}^\pm$ of $\CZ^{\bf n}$, 
\bea(lll)
{\cal C}^+ = \sum_{\sigma =1}^{N-1} \CZ |\sigma \rangle \rangle_+ = \sum_{k=1}^{N-1} \CZ  \widehat{|k} \rangle \rangle_+ , &   \widehat{|k} \rangle \rangle_+ := \frac{1}{2} (\widehat{|k } \rangle \rangle + \widehat{|k +N} \rangle \rangle),  \\
{\cal C}^- = \sum_{\sigma =1}^{N-1} \CZ |\sigma \rangle \rangle_-  = \sum_{k=1}^{N-1} \CZ \widehat{|k} \rangle \rangle_-, & \widehat{|k} \rangle \rangle_- := \frac{1}{2} (q^k\widehat{|k } \rangle \rangle +q^{k+N} \widehat{|k +N} \rangle \rangle), 
\elea(Cpm)
where $\widehat{|k } \rangle \rangle_\pm$'s are basis of ${\cal C}^\pm$, related to  $|\sigma \rangle \rangle_\pm $'s by the $N$-Fourier relation:
$$
\begin{array}{ll}
\widehat{|k } \rangle \rangle_\pm  = \frac{1}{\sqrt{N}} \sum_{\sigma =0}^{N-1} \omega^{-k \sigma} |\sigma \rangle \rangle_\pm , & | \sigma  \rangle \rangle_\pm = \frac{1}{\sqrt{N}} \sum_{k =0}^{N-1} \omega^{\sigma k} \widehat{|k} \rangle \rangle_\pm , ~ (\sigma  \in \ZZ_N). 
\end{array}
$$
Then 

(i) $ \wp_+ = \wp_- \widehat{Z}^\prime$, and $ \widehat{Z}^{-1} \wp_- = \wp_+ \widehat{Z}^\prime$.

(ii) The ${\cal C}^\pm$-bases  in (\req(Cpm)) interchange under $\CZ^{\bf n}$-Weyl operators via
\bea(llll)
X' : {}^{|\sigma \rangle \rangle_+}_{|\sigma \rangle \rangle_-} \mapsto {}^{|\sigma \rangle \rangle_-}_{|\sigma - 1 \rangle \rangle_+},&
Z' : {}^{|\sigma \rangle \rangle_+}_{|\sigma \rangle \rangle_-}  \mapsto {}^{\omega^\sigma |\sigma \rangle \rangle_+}_{q \omega^k|\sigma + 1 \rangle \rangle_- }  ;&
\widehat{X}': {}^{\widehat{|k } \rangle \rangle_+}_{\widehat{|k } \rangle \rangle_- } \mapsto {}^{\widehat{|k +1} \rangle \rangle_+}_{q^{-1} \widehat{|k+1 } \rangle_-} ,&
\widehat{Z}': {}^{\widehat{|k } \rangle \rangle_+}_{\widehat{|k } \rangle \rangle_- } \mapsto {}^{\widehat{|k } \rangle \rangle_-}_{ \omega^{-k} \widehat{|k }\rangle \rangle_+}, 
\elea(XZpm)
by which $\CZ^N$-Weyl operators of ${\cal C}^\pm$ are induced by
\bea(llllll)
(X^{'-2}, Z'), & (\widehat{X}' , \widehat{Z}^{' -2}) & {\rm for} ~ ~ {\cal C}^+ ; &
 (X^{'-2}, q^{-1}Z'),& (q \widehat{X}', \widehat{Z}^{' -2}) & {\rm for} ~ ~ {\cal C}^-.
\elea(WCpm)
 Furthermore, the $\CZ^{\bf n}$-operator $X'$ identifies the $N$-spaces ${\cal C}^\pm$ by 
\bea(lll)
X' (= \widehat{Z}') : {\cal C}^+ \simeq {\cal C}^- , & |\sigma \rangle \rangle_+  \mapsto |\sigma \rangle \rangle_- , & \widehat{|k } \rangle \rangle_+ \mapsto \widehat{|k } \rangle \rangle_- ,
\elea(Cpm=)
and ${\cal C}^\pm$ are isomorphic to $\CZ^N$ under $\wp_\pm$ in (\req(wp)) respectively:
\bea(lll)
\wp_\pm : {\cal C}^\pm \simeq \CZ^N, & |\sigma \rangle \rangle_\pm  \mapsto |\sigma \rangle, & \widehat{|k } \rangle \rangle_\pm \mapsto \widehat{|k } \rangle .
\elea(CpmCN)
The automorphism (\req(Cpm=)) descends to the following $\CZ^N$-automorphisms via $\wp_\pm$ in (\req(CpmCN)):
\bea(llll)
\CZ^N \stackrel{\wp_+}{\simeq } {\cal C}^+ \stackrel{\widehat{Z}'}{\simeq } {\cal C}^- \stackrel{\wp_-}{\simeq } \CZ^N , 
& \widehat{|k } \rangle \mapsto \widehat{|k } \rangle ; &
\CZ^N \stackrel{\wp_-}{\simeq } {\cal C}^- \stackrel{\widehat{Z}'}{\simeq } {\cal C}^+ \stackrel{\wp_+}{\simeq } \CZ^N , 
& \widehat{|k } \rangle \mapsto \omega^{-k} \widehat{|k } \rangle .
\elea(gwpN)

(iii) When ${\bf n}=N$ odd, $\CZ^{\bf n}= {\cal C}^\pm$ with $(|\sigma \rangle \rangle_-,  \widehat{|k} \rangle \rangle_- ) = (|\sigma + \frac{N-1}{2} \rangle \rangle_+,  q^k \widehat{|k} \rangle \rangle_+)$, and the projections (\req(CpmCN)) define isomorphisms between $\CZ^{\bf n}$ and $\CZ^N$, preserving the Fourier basis. When  ${\bf n}=2N$, $\CZ^{\bf n} = {\cal C}^+ \oplus {\cal C}^-$, and the kernel of (\req(wp)) $ {\rm Ker} (\wp_\mp )= {\cal C}^\pm$. Hence $|\sigma \rangle \rangle_\pm, \widehat{|k } \rangle \rangle_\pm$'s form a basis of $\CZ^{\bf n}$.  In both cases, the Weyl operators of $\CZ^{\bf n}$ and $\CZ^N$ are related by 
\bea(lll)
 \wp_\pm  X^{' -2} = X \wp_\pm   , q^\frac{-1 \pm 1}{2}  \wp_\pm Z'=  Z \wp_\pm  , & ( \Leftrightarrow &  q^\frac{1 \mp 1}{2} \wp_\pm \widehat{X}' =  \widehat{X} \wp_\pm ,   \wp_\pm \widehat{Z}^{' -2}= \widehat{Z} \wp_\pm ) .  
\elea(Wid)
\end{lem}
$\Box$ \par \vspace{.1in} \noindent
We shall also consider another kind of cyclic $N$-subspaces of $\CZ^{\bf n}$.
\begin{lem}\label{lem:CnNdag} Consider the cyclic $N$-vectors in $\CZ^{\bf n}$,
$$
 \widehat{|k} \rangle \rangle^\dagger_+ := c({\bf n}) \widehat{|-2 k} \rangle \rangle, ~ \widehat{|k} \rangle \rangle^\dagger_- := c({\bf n}) \widehat{|-2 k +1} \rangle \rangle \in \CZ^{\bf n} ~  ~ ~ (k \in \ZZ_N),
$$
and the linear transformations
\bea(lll)
\wp^\dagger_\pm : \CZ^{\bf n} \longrightarrow \CZ^N , ~ & \wp^\dagger_+( |\sigma \rangle \rangle )= |\sigma \rangle , &
\wp^\dagger_-( |\sigma \rangle \rangle )= q^\sigma |\sigma \rangle ~ ~ ( \sigma \in \ZZ_{\bf n}).
\elea(wpdag)
Define the  cyclic $N$-subspaces ${\cal C}^{\dagger \pm}$ of $\CZ^{\bf n}$, 
\bea(lll)
{\cal C}^{\dagger +} = \sum_{\sigma =1}^{N-1} \CZ |\sigma \rangle \rangle^\dagger_+ = \sum_{k=1}^{N-1} \CZ  \widehat{|k} \rangle \rangle^\dagger_+ , &   |\sigma \rangle \rangle^\dagger_+ := \frac{1}{2} (|\sigma \rangle \rangle + |\sigma +N \rangle \rangle),  \\
{\cal C}^{\dagger -} = \sum_{\sigma =1}^{N-1} \CZ |\sigma \rangle \rangle^\dagger_-  = \sum_{k=1}^{N-1} \CZ \widehat{|k} \rangle \rangle^\dagger_-, & |\sigma \rangle \rangle^\dagger_- := \frac{1}{2} (q^{-\sigma} |\sigma \rangle \rangle +q^{-\sigma-N} |\sigma +N \rangle \rangle), 
\elea(Cpmdag)
with the relation of basis:
$$
\begin{array}{ll}
\widehat{|k } \rangle \rangle^\dagger_\pm  = \frac{1}{\sqrt{N}} \sum_{\sigma =0}^{N-1} \omega^{-k \sigma} |\sigma \rangle \rangle^\dagger_\pm , & | \sigma  \rangle \rangle^\dagger_\pm = \frac{1}{\sqrt{N}} \sum_{k =0}^{N-1} \omega^{\sigma k} \widehat{|k} \rangle \rangle^\dagger_\pm , ~ (\sigma  \in \ZZ_N). 
\end{array}
$$
Then 

(i) $ \wp^\dagger_+ = \wp^\dagger_- Z^{\prime -1}$, and $ Z \wp^\dagger_- = \wp^\dagger_+ Z^{\prime -1}$.

(ii) The $\CZ^{\bf n}$-Weyl operators interchange the  ${\cal C}^{\dagger \pm}$-basis in (\req(Cpmdag)) by  
\bea(llll)
X' : {}^{|\sigma \rangle \rangle^\dagger_+}_{|\sigma \rangle \rangle^\dagger_-} \mapsto {}^{|\sigma +1 \rangle \rangle^\dagger_+}_{q|\sigma + 1 \rangle \rangle^\dagger_-},&
Z' : {}^{|\sigma \rangle \rangle^\dagger_+}_{|\sigma \rangle \rangle^\dagger_-}  \mapsto {}^{\omega^{-\sigma} |\sigma \rangle \rangle^\dagger_-}_{|\sigma \rangle \rangle^\dagger_+ }  ;&
\widehat{X}': {}^{\widehat{|k } \rangle \rangle^\dagger_+}_{\widehat{|k } \rangle \rangle^\dagger_- } \mapsto {}^{\widehat{|k} \rangle \rangle^\dagger_-}_{\widehat{|k-1 } \rangle^\dagger_+} ,& 
\widehat{Z}': {}^{\widehat{|k } \rangle \rangle^\dagger_+}_{\widehat{|k } \rangle \rangle^\dagger_- } \mapsto {}^{\omega^k \widehat{|k } \rangle \rangle^\dagger_+}_{ q \omega^k \widehat{|k }\rangle \rangle^\dagger_-}, 
\elea(XZpmdag)
by which the $\CZ^N$-Weyl operators of ${\cal C}^{\dagger \pm}$ are induced by 
\bea(llllll)
(X', Z^{\prime -2}), & (\widehat{X}^{' -2} , \widehat{Z}') & {\rm for} ~ ~ {\cal C}^{\dagger +} ; &
 (q^{-1} X', Z^{\prime -2}),& (\widehat{X}^{' -2}, q^{-1} \widehat{Z}') & {\rm for} ~ ~ {\cal C}^{\dagger -}.
\elea(WCpmdag)
Furthermore, the $\CZ^{\bf n}$-operator $Z^{\prime -1}$ identifies the $N$-spaces ${\cal C}^{\dagger \pm}$:
\bea(lll)
Z^{\prime -1} (= \widehat{X}') : {\cal C}^{\dagger +} \simeq {\cal C}^{\dagger -} , & |\sigma \rangle \rangle^\dagger_+  \mapsto |\sigma \rangle \rangle^\dagger_- , & \widehat{|k } \rangle \rangle^\dagger_+ \mapsto \widehat{|k } \rangle \rangle^\dagger_- ,
\elea(Cpm=dag)
and ${\cal C}^{\dagger \pm}$ are isomorphic to $\CZ^N$ under $\wp^\dagger_\pm$ in (\req(wpdag)) respectively:
\bea(lll)
\wp^\dagger_\pm : {\cal C}^{\dagger \pm} \simeq \CZ^N, & |\sigma \rangle \rangle^\dagger_\pm  \mapsto |\sigma \rangle, & \widehat{|k } \rangle \rangle^\dagger_\pm \mapsto \widehat{|k } \rangle .
\elea(CdagCN)
The automorphism (\req(Cpm=dag)) descends to the following $\CZ^N$-automorphisms via $\wp^\dagger_\pm$ in (\req(CdagCN)):
\bea(llll)
\CZ^N \stackrel{\wp^\dagger_+}{\simeq } {\cal C}^{\dagger +} \stackrel{\widehat{X}'}{\simeq } {\cal C}^{\dagger -} \stackrel{\wp^\dagger_-}{\simeq } \CZ^N , 
& |\sigma \rangle \mapsto |\sigma \rangle ; &
\CZ^N \stackrel{\wp^\dagger_-}{\simeq } {\cal C}^{\dagger -} \stackrel{\widehat{X}'}{\simeq } {\cal C}^{\dagger +} \stackrel{\wp^\dagger_+}{\simeq } \CZ^N , 
& |\sigma \rangle \mapsto \omega^\sigma |\sigma \rangle .
\elea(gwpN)

(iii) When ${\bf n}=N$ odd, $\CZ^{\bf n}= {\cal C}^{\dagger \pm}$ with $(|\sigma \rangle \rangle^\dagger_-,  \widehat{|k} \rangle \rangle^\dagger_- ) = (q^{-\sigma} |\sigma \rangle \rangle^\dagger_+,  \widehat{|k+ \frac{N-1}{2}} \rangle \rangle^\dagger_+)$, and (\req(CdagCN)) defines the isomorphism between $\CZ^{\bf n}$ and $\CZ^N$, preserving the Fourier basis. When  ${\bf n}=2N$, $\CZ^{\bf n} = {\cal C}^{\dagger +} \oplus {\cal C}^{\dagger -}$ with ${\cal C}^{\dagger \pm} = {\rm Ker} (\wp^\dagger_\mp )$, and  $|\sigma \rangle \rangle^\dagger_\pm, \widehat{|k } \rangle \rangle^\dagger_\pm$'s form a basis of $\CZ^{\bf n}$.  In both cases, the Weyl operators of $\CZ^{\bf n}$ and $\CZ^N$ are related by 
\bea(lll)
 \wp^\dagger_\pm  X' = q^\frac{1 \mp 1}{2} X \wp^\dagger_\pm , \wp^\dagger_\pm Z^{' -2}=  Z \wp^\dagger_\pm  , & ( \Leftrightarrow &   \wp^\dagger_\pm \widehat{X}^{' -2} =  \widehat{X} \wp^\dagger_\pm ,   q^\frac{-1 \pm 1}{2}\wp^\dagger_\pm \widehat{Z}'= \widehat{Z} \wp^\dagger_\pm ) .  
\elea(Widdag)
\end{lem}
$\Box$ \par \vspace{.1in} \noindent

\subsection{The $L$-operators of  XXZ-model and  $\tau^{(2)}$-model with cyclic representation  \label{ssec.L}}
Consider the cyclic-representation of $U_q ( sl_2)$, i.e. the three-parameter family of $U_q ( sl_2)$-representation on the cyclic space $\CZ^{\bf n}$ with parameters $q^{\phi}, q^{-\phi^\prime}$ and $q^{\varepsilon}$, denoted by $s_{\varepsilon, \phi, \phi^\prime}$, and  with the following expression in terms of $\CZ^{\bf n}$-Weyl operators:
\bea(lll)
K^\frac{1}{2} = q^\frac{\phi^\prime- \phi}{2}  \widehat{Z}',& 
e^+  = q^{ \varepsilon} \frac{(  q^{\phi +1} \widehat{Z}^{\prime -1}  -  q^{- \phi -1}\widehat{Z}')\widehat{X}^\prime}{q-q^{-1}} &
e^-  =q^{ -\varepsilon} \frac{(  q^{\phi^\prime +1} \widehat{Z}' -  q^{- \phi^\prime -1} \widehat{Z}^{\prime -1})\widehat{X}^{\prime -1}}{q-q^{-1}},
\elea(crep)
or equivalently, a expression using the spin-operators $(X', Z')= (\widehat{Z}', \widehat{X}^{' -1})$ (see, e.g. \cite{DJMM, DK} or \cite{R0806, R09}).  By (\req(6vL)), the $U_q ( sl_2)$-representation (\req(crep)) is equivalent to the YB relation (\req(YBXXZ)) of the $L$-operator of XXZ-model:
\bea(lc)
{\cal L} (s)   = & \left( \begin{array}{cc}
         s \rho^{-1} \nu^\frac{1}{2} q^\frac{- \phi^\prime + \phi}{2}  \widehat{Z}^{' -1}  -  s^{-1} \nu^\frac{-1}{2} q^\frac{\phi^\prime- \phi}{2}  \widehat{Z}'   &  q^{ -\varepsilon} (  q^{\phi^\prime +1} \widehat{Z}' -  q^{- \phi^\prime -1} \widehat{Z}^{\prime -1})\widehat{X}^{\prime -1}    \\
        q^{ \varepsilon} (  q^{\phi +1} \widehat{Z}^{\prime -1}  -  q^{- \phi -1}\widehat{Z}')\widehat{X}^\prime & s \nu^\frac{1}{2} q^\frac{\phi^\prime- \phi}{2}  \widehat{Z}' -  s^{-1} \rho \nu^\frac{-1}{2} q^\frac{-\phi^\prime+ \phi}{2}  \widehat{Z}^{' -1}
\end{array} \right) .
\elea(6Lcy)
The above representations are all irreducible except a special one in ${\bf n}=2N$ case which is equivalent to the spin-$\frac{{\bf n}-1}{2}$ representation in (\req(spinrp)). The $U_q ( sl_2)$-representation (\req(crep)) induces the $\CZ^{\bf n}$-representation of ${\Large\textsl{U}}_\omega (sl_2)$ where the generators in (\req(qUw)) are expressed by 
\bea(lll)
K = q^{\phi^\prime - \phi } \widehat{Z}^{\prime 2} , &  
E^+ = q^{\frac{-\phi - \phi^\prime}{2} + \varepsilon} \frac{( 1 -\omega^{-\phi-1} \widehat{Z}^{\prime - 2} )\widehat{X}^\prime }{1- \omega }, &
E^-  = q^{\frac{\phi+\phi^\prime}{2} - \varepsilon} \frac{(  1   - \omega^{\phi^\prime +1} \widehat{Z}^{\prime -2} )\widehat{X}^{\prime -1}}{1-\omega }.  
\elea(crepU) 
The $L$-operator (\req(Ltau)) with the $\CZ^{\bf n}$-representation (\req(crepU)),
\bea(lc)
 \textsc{L} (t) & = \left( \begin{array}{cc}
        1- t \nu  \rho^{-1} q^{\phi - \phi^\prime } \widehat{Z}^{\prime -2}  &  q^{\frac{\phi+\phi^\prime}{2} - \varepsilon} (  1   - \omega^{\phi^\prime +1} \widehat{Z}^{\prime -2} )\widehat{X}^{\prime -1}    \\
         - t \nu q^{\frac{-\phi - \phi^\prime}{2} + \varepsilon} ( 1 -\omega^{-\phi-1} \widehat{Z}^{\prime - 2} )\widehat{X}^\prime  & - t \nu   +    \rho q^{\phi -\phi^\prime  } \widehat{Z}^{\prime - 2}
\end{array} \right),
\elea(LtCn)
then satisfies the YB relation (\req(YBt2)). By (\req(Wid)), representations in (\req(crepU)) descend to the following three-parameter family of cyclic $\CZ^N$-representation of ${\Large\textsl{U}}_\omega (sl_2)$ via $\wp_+$ in (\req(wp)):
\bea(lll)
K = q^{\phi^\prime - \phi }\widehat{Z}^{-1} , &  
E^+ = q^{\frac{-\phi - \phi^\prime}{2} + \varepsilon} \frac{( 1 -\omega^{-\phi-1} \widehat{Z} )\widehat{X} }{1- \omega }, &
E^-  = q^{\frac{\phi+\phi^\prime}{2} - \varepsilon} \frac{(  1   - \omega^{\phi^\prime +1} \widehat{Z} )\widehat{X}^{ -1}}{1-\omega }. 
\elea(crXZ) 
By employing the cyclic representation (\req(crXZ)) of ${\Large\textsl{U}}_\omega (sl_2)$ on the two-parameter family of $L$-operator (\req(Ltau)), one obtains the following five-parameter $L$-operators of the cyclic $\CZ^N$-space appeared in $\tau^{(2)}$-model of the $N$-state CPM \cite{BazS}:
\bea(lll)
\textsl{L}(t) 
& = \left( \begin{array}{cc}
        1- t \frac{\sf c}{\sf b'b } \widehat{Z}  &   ( \frac{1}{\sf b }   - \omega \frac{\sf a c}{\sf b'b } \widehat{Z}  )\widehat{X}^{-1}    \\
         - t     ( \frac{1}{\sf b'} -\frac{{\sf a' c}}{\sf b'b } \widehat{Z}  )\widehat{X} & - t \frac{1}{\sf b'b }   +   \omega \frac{ {\sf a'a c}}{\sf b' b } \widehat{Z} 
\end{array} \right), & (\widehat{Z}, \widehat{X}) = (X, Z^{-1}) ,
\elea(LtauC)
where as  in \cite{R09}, the parameters ${\sf a', b', a, b, c}$ are related to $\varepsilon, \phi, \phi^\prime, \rho, \nu$ by\footnote{Here the parameter $({\sf a', b', a, b, c})$ is equal to $(\nu^\frac{1}{2} {\sf a}', \nu^\frac{-1}{2} {\sf b}', \nu^\frac{-1}{2} {\sf a}, \nu^\frac{1}{2} {\sf b}, {\sf c})$ in formula (4.17) of \cite{R09}. The difference is due to $L(t)$ in \cite{R09} is identified with the gauge of $-s \nu^\frac{1}{2} K^{\frac{-1}{2}} {\cal L} (s)$ by ${\rm dia}[1, -sq]$, instead of ${\rm dia}[1, -s\nu^\frac{1}{2}q]$ in this paper.} :
\bea(l)
 q^{\varepsilon + \frac{1}{2}} = (\frac{{\sf a}'{\sf b}'{\sf b}^3}{{\sf a} })^\frac{1}{4} , ~ q^{\phi+1}= (\frac{{\sf a' c}}{\sf b})^\frac{1}{2} , ~ q^{- \phi^\prime}= (\frac{\sf a c}{\sf b'})^\frac{1}{2} , ~ \rho = q^{-1}(\frac{{\sf a'a }}{\sf b'b })^\frac{1}{2}, \nu =  \frac{1}{\sf b' b} ~ ~ \Longleftrightarrow   \\
{\sf a} = \rho \nu^{-1}  q^{\frac{-\phi - \phi^\prime}{2} - \varepsilon} , ~ \omega {\sf a'}{\sf a}  = \rho^2 \nu^{-1}, ~  {\sf b}  = q^{\frac{-\phi - \phi^\prime}{2}+ \varepsilon},  ~   {\sf b'}{\sf b } = \nu^{-1},  ~ {\sf c}  = \rho^{-1} q^{\phi - \phi^\prime} . 
\elea(par=)
Note that representations in (\req(crXZ)), when changing $q^\varepsilon$ to $\omega^{-n} q^\varepsilon$, or equivalently, $( {\sf a'},  {\sf b'}, {\sf a}, {\sf b}, {\sf c})$ to $( \omega^{-n}{\sf a'}, \omega^n {\sf b'}, \omega^n {\sf a}, \omega^{-n}{\sf b}, {\sf c})$ for $n \in \ZZ$,  are equivalent. For convenience, through a factorization of ${\sf c}$, we shall express the parameter of $L$-operator in (\req(LtauC)) by 
\bea(llll)
\textsl{L}(t; {\sf p}', {\sf p}) := \textsl{L}(t), &{\sf p}' = ({\sf a', b', d'}), & {\sf p}= ({\sf a, b, d}), & {\sf c}={\sf d' d}.
\elea(Lp'p) 
We shall also write ${\sf p}_+' :={\sf p}', {\sf p}_+ :={\sf p}$. Through $\wp_-$ in (\req(wp)) and relations in (\req(Wid)), the ${\Large\textsl{U}}_\omega (sl_2)$-representation (\req(crepU))  and $L$-operator (\req(LtCn)) descend to the cyclic $\CZ^N$-representation (\req(crXZ)) with $\varepsilon$ replaced by $\varepsilon -1$, hence the $L$-operator (\req(LtauC)) with parameter $({\sf p}_-', {\sf p}_-)$,
\bea(lll) 
\textsl{L}(t; {\sf p}_-', {\sf p}_-) , & {\sf p}_-' = ({\sf a'_-, b'_-, d'_-}), & {\sf p}_- = ({\sf a_-, b_-, d_-}),
\elea(Lp'p-)
where $({\sf p}_-', {\sf p}_-)$ is related to  $({\sf p'}, {\sf p})$ by 
\bea(lll)
({\sf  a_-', b_-', a_-, b_-, c_-})= ( q^{-1}{\sf a'}, q {\sf b'}, q{\sf a}, q^{-1}{\sf b}, {\sf c}),&  c_-= {\sf d}_-'{\sf d}_-, & ({\sf d}_-', {\sf d}_-) = ({\sf d}', {\sf d}). 
\elea(aminus)
The relation (\req(aminus)) is equivalent to the gauge-equivalence of $L$-operators:
\bea(cll) 
\textsl{L}(t; {\sf p}_-', {\sf p}_-)&=  {\rm dia}[1, q^{-1}]\textsl{L}(t; {\sf p}_+'{\sf p}_+ ){\rm dia}[1, q], & \Leftrightarrow \\
  \widehat{Z} \textsl{L}(t; {\sf p}_+', {\sf p}_+ ) \widehat{Z}^{-1}& =  {\rm dia}[1, q^{-1}] \textsl{L}(t; {\sf p}_-', {\sf p}_-){\rm dia}[1, q]. 
\elea(Lgaequ)
The $L$-operators in (\req(Lp'p)), (\req(Lp'p-)) are related to the $\CZ^{\bf n}$-representation in (\req(crepU)) in the following result in \cite{R0806}:
\begin{lem}\label{lem:Cpm} The $N$-subspaces ${\cal C}^\pm$ of $\CZ^{\bf n}$ in (\req(Cpm)) are irreducible components of the ${\Large\textsl{U}}_\omega (sl_2)$-representation (\req(crepU)) defined by the Weyl operators in (\req(WCpm)). As ${\Large\textsl{U}}_\omega (sl_2)$-representations, ${\cal C}^\pm$ is equivalent to $\textsl{L}({\sf p}_\pm', {\sf p}_\pm; t)$ via the morphism $\wp_\pm$ in (\req(CpmCN)). For ${\bf n}=N$ odd, $\CZ^{\bf n} \simeq {\cal C}^\pm$ with the equivalence between ${\cal C}^\pm$ given by $\widehat{|k} \rangle  \rangle_-= q^k \widehat{|k} \rangle  \rangle_+$.  For ${\bf n}=2N$, $\CZ^{\bf n} = {\cal C}^+ \oplus {\cal C}^-$, which is equivalent to $\textsl{L}(t; {\sf p}_+', {\sf p}_+) \oplus \textsl{L}(t; {\sf p}_-', {\sf p}_-)$, as ${\Large\textsl{U}}_\omega (sl_2)$-representations.
\end{lem}
{\it Proof.} It is obvious that ${\cal C}^\pm$ with the Weyl operators in (\req(WCpm)) are irreducible components of ${\Large\textsl{U}}_\omega (sl_2)$-representation $\CZ^{\bf n}$ in (\req(crepU)). The rest statements follows from the relations 
(\req(CpmCN)), (\req(crXZ))-(\req(Lp'p-)), and  Lemma \ref{lem:CnN} $(iii)$. 
$\Box$ \par \vspace{.1in} \noindent
Through the substitution (\req(par=)), we shall also use the parameter $({\sf p}', {\sf p})$ in (\req(Lp'p)) to represent the $L$-operators ${\cal L} (s)$, $\textsc{L} (t)$  in (\req(6Lcy)) and (\req(LtCn)): 
\bea(ll)
{\cal L} (s; {\sf p}', {\sf p})   = & (\frac{\sf  b^{' 3}b^3}{\omega {\sf a'a c^2}})^\frac{1}{4}  \left( \begin{array}{cc}
          - s^{-1}  \widehat{Z}' + s \frac{\sf c}{\sf b'b} \widehat{Z}^{' -1}     &  \frac{q}{({\sf b' b})^\frac{1}{2} }
 (  \frac{1}{\sf b} \widehat{Z}' - \omega \frac{\sf ac}{\sf b'b}  \widehat{Z}^{\prime -1})\widehat{X}^{\prime -1}    \\
      \frac{-({\sf b' b})^\frac{1}{2}}{q}  ( \frac{1}{\sf b'} \widehat{Z}'  -\frac{{\sf a' c}}{\sf b' b} \widehat{Z}^{\prime -1}  )\widehat{X}^\prime & s  \frac{1}{\sf b' b}   \widehat{Z}' -  s^{-1} \omega \frac{{\sf a'a c}}{\sf b'b }  \widehat{Z}^{' -1}
\end{array} \right) , \\
\textsc{L} (t; {\sf p}', {\sf p})= & \left( \begin{array}{cc}
        1- t \frac{\sf c}{\sf b'b} \widehat{Z}^{\prime -2}  & (   \frac{1}{\sf b}    - \omega \frac{\sf ac}{\sf b'b} \widehat{Z}^{\prime -2} )\widehat{X}^{\prime -1}    \\
         - t  (  \frac{1}{\sf b'}  -\frac{{\sf a' c}}{\sf b' b}\widehat{Z}^{\prime - 2} )\widehat{X}^\prime  & - t \frac{1}{\sf b' b}  +    \omega \frac{{\sf a'a c}}{\sf b'b }  \widehat{Z}^{\prime - 2}
\end{array} \right) , ~  ~ (\widehat{Z}', \widehat{X}') = (X', Z^{' -1}),
\elea(LLp'p)
which are related by
\bea(ll)
\textsc{L} (t; {\sf p}'{\sf p} ) = {\rm dia}[1, - s({\sf b'}{\sf b })^\frac{-1}{2} q]\bigg(-s (\frac{\omega {\sf a'a c^2}}{\sf  b^{' 3}b^3})^\frac{1}{4}\widehat{Z}^{'-1}  {\cal L} (s; {\sf p}'{\sf p} ) \bigg) {\rm dia}[1, -s^{-1}({\sf b'}{\sf b })^\frac{1}{2}q^{-1}]  \Leftrightarrow \\
{\cal L} (s; {\sf p}'{\sf p} )  = {\rm dia}[1, - s^{-1}({\sf b'}{\sf b })^\frac{1}{2} q^{-1}] \bigg(-s^{-1} (\frac{\omega {\sf a'a c^2}}{\sf  b^{' 3}b^3})^\frac{-1}{4}\widehat{Z}' \textsc{L} (t; {\sf p}'{\sf p} ) \bigg)  {\rm dia}[1, -s({\sf b'}{\sf b })^\frac{-1}{2}q] .  \\
\elea(LLrel)
Corresponding to (\req(Lgaequ)), we have
\bea(lll) 
{\cal L}(s; {\sf p}_-', {\sf p}_-)=  \widehat{Z}^{' -1}{\cal L}(s; {\sf p}_+'{\sf p}_+ )\widehat{Z}', & (\Leftrightarrow &
  {\cal L}(s; {\sf p}_+', {\sf p}_+ )  =  \widehat{Z}' {\cal L}(s; {\sf p}_-', {\sf p}_-)\widehat{Z}^{' -1}) ; \\
\textsc{L} (t; {\sf p}_-', {\sf p}_-)=  \widehat{Z}^{' -1} \textsc{L} (t; {\sf p}_+'{\sf p}_+ )\widehat{Z}', & (\Leftrightarrow &
 \textsc{L} (t; {\sf p}_+', {\sf p}_+ )  =  \widehat{Z}' \textsc{L} (t; {\sf p}_-', {\sf p}_-)\widehat{Z}^{' -1}). 
\elea(LXXZg)
Indeed, by (\req(gwpN)), the second relation in (\req(LXXZg)) is equivalent to the gauge relations in (\req(Lgaequ)).  

There is another type of cyclic $\CZ^{\bf n}$-representations of $(\req(Ltau))_{\textsl{w}=\omega}$ associated to the following ${\Large\textsl{U}}_\omega (sl_2)$ representation:
\bea(lll)
K = q^{\phi^\prime - \phi } \widehat{Z}^{\prime -1} , &  
E^+ = q^{\frac{-\phi - \phi^\prime}{2} + \varepsilon} \frac{( 1 -\omega^{-\phi-1} \widehat{Z}^{\prime} )\widehat{X}^{\prime -2} }{1- \omega }, &
E^-  = q^{\frac{\phi+\phi^\prime}{2} - \varepsilon} \frac{(  1   - \omega^{\phi^\prime +1} \widehat{Z}^{\prime } )\widehat{X}^{\prime 2}}{1-\omega }, 
\elea(cycUdag)
with the $L$-operator
\bea(ll)
\textsc{L}^\dagger (t; {\sf p}', {\sf p})= & \left( \begin{array}{cc}
        1- t \frac{\sf c}{\sf b'b} \widehat{Z}^\prime   & (   \frac{1}{\sf b}    - \omega \frac{\sf ac}{\sf b'b} \widehat{Z}^\prime  )\widehat{X}^{\prime 2}    \\
         - t  (  \frac{1}{\sf b'}  -\frac{{\sf a' c}}{\sf b' b}\widehat{Z}^\prime )\widehat{X}^{\prime -2} & - t \frac{1}{\sf b' b}  +    \omega \frac{{\sf a'a c}}{\sf b'b }  \widehat{Z}^\prime
\end{array} \right) , ~  ~ (\widehat{Z}', \widehat{X}') = (X', Z^{' -1}).
\elea(Lp'pdag)
The representation (\req(cycUdag)) and $L$-operator (\req(Lp'pdag)) are obtained by the replacement of $(\widehat{X}^\prime , \widehat{Z}^{\prime - 2})$ in (\req(crepU)), (\req(LLp'p)) by $(\widehat{X}^{\prime -2}, \widehat{Z}^\prime)$. 
By (\req(Widdag)) and (\req(par=)), the representation (\req(cycUdag)) and $L$-operator (\req(Lp'pdag)) descend to the $\CZ^N$-representation (\req(crXZ)) with the parameter $(\varepsilon^\dagger_\pm, \phi^\dagger_\pm, \phi^{\prime \dagger}_\pm)$  and $\textsl{L}(t; {\sf p}^{\prime \dagger}_\pm , {\sf p}^\dagger_\pm)$ in (\req(LtauC)), via $\wp_\pm^\dagger$ in (\req(wpdag)), where the parameters are defined by
\bea(ll)
(\varepsilon^\dagger_+, \phi^\dagger_+, \phi^{\prime \dagger}_+) = (\varepsilon, \phi, \phi^{\prime}), 
 & {\sf p}^{\prime \dagger}_+ = {\sf p}^{\prime}= ({\sf a}^\prime, {\sf b}^\prime, {\sf d}^\prime) , {\sf p}^\dagger_+= {\sf p}= ({\sf a}, {\sf b}, {\sf d}) ; \\
(\varepsilon^\dagger_-, \phi^\dagger_-, \phi^{\prime \dagger}_-)=(\varepsilon, \phi+\frac{1}{2},\phi'-\frac{1}{2}), & {\sf p}^{\prime \dagger}_-= ({\sf a}^\prime , {\sf b}^\prime, {\sf d}^\prime) , {\sf p}^\dagger_-= ({\sf a}, {\sf b}, {\sf d}q) .
\elea(pp'dag)
As in (\req(LXXZg)), we have 
\bea(lll) 
\textsc{L}^\dagger (t; {\sf p}_-^{\prime \dagger }, {\sf p}^\dagger_-)=  \widehat{X}^{' -1} \textsc{L}^\dagger (t; {\sf p}_+^{\prime \dagger}{\sf p}^\dagger_+ )\widehat{X}' , & (\Leftrightarrow &
 \textsc{L}^\dagger (t; {\sf p}_+^{' \dagger}, {\sf p}_+^\dagger )  =  \widehat{X}' \textsc{L}^\dagger (t; {\sf p}_-^{\prime \dagger}, {\sf p}^\dagger_-)\widehat{X}^{\prime -1}), 
\elea(LXXZg)
which by (\req(gwpN)), is equivalent to 
$$
\begin{array}{ll} 
\textsl{L}(t; {\sf p}^{' \dagger}_-, {\sf p}^\dagger_-) =  \widehat{X}^\frac{1}{2}\textsl{L}(t; {\sf p}^{' \dagger}_+{\sf p}^\dagger_+ )\widehat{X}^\frac{-1}{2} , &
 \widehat{X} \textsl{L}(t; {\sf p}^{' \dagger}_+, {\sf p}^\dagger_+ ) \widehat{X}^{-1} =  \widehat{X}^\frac{1}{2} \textsl{L}(t; {\sf p}^{' \dagger}_-, {\sf p}^\dagger_-)\widehat{X}^\frac{-1}{2}, 
\end{array}
$$
where $\widehat{X}^\frac{-1}{2} := \widehat{X}'$. Note that the above relation is similar to that in (\req(Lgaequ)), but not in a form of gauge equivalence. As in Lemma \ref{lem:Cpm}, we have the following result:
\begin{lem}\label{lem:Cdagpm} The $N$-subspaces ${\cal C}^{\dagger \pm}$ of $\CZ^{\bf n}$ in (\req(Cpmdag)) are irreducible components of the ${\Large\textsl{U}}_\omega (sl_2)$-representation (\req(cycUdag)). As ${\Large\textsl{U}}_\omega (sl_2)$-representations, ${\cal C}^{\dagger \pm}$ is equivalent to $\textsl{L}({\sf p}^{\prime \dagger}_\pm, {\sf p}^\dagger_\pm; t)$ via the morphism $\wp^\dagger_\pm$ in (\req(CdagCN)). For ${\bf n}=N$ odd, $\CZ^{\bf n} \simeq {\cal C}^{\dagger \pm}$ with the equivalence between ${\cal C}^{\dagger \pm}$ given by $|\sigma \rangle \rangle^\dagger_- = q^{-\sigma} |\sigma \rangle\rangle^\dagger_+$. For ${\bf n}=2N$, $\CZ^{\bf n} = {\cal C}^{\dagger +} \oplus {\cal C}^{\dagger -}$ is equivalent to $\textsl{L}(t; {\sf p}^{\prime \dagger}_+, {\sf p}^\dagger_+) \oplus \textsl{L}(t; {\sf p}^{\prime \dagger}_-, {\sf p}^\dagger_-)$ as ${\Large\textsl{U}}_\omega (sl_2)$-representations.
\end{lem}
$\Box$ \par \vspace{.1in} \noindent
The following lemma describes the equivalent representations in (\req(crep)), (\req(crepU) ) or (\req(cycUdag)) under the change of  parameter $({\sf p}', {\sf p})$.
\begin{lem}\label{lem:l-p} We define the $l$-twists of ${\sf p}=({\sf a}, {\sf b}, {\sf d})$ by 
\bea(lll)
{\sf p}(l):= ({\sf a}q^l, {\sf b}q^{-l}, {\sf d}),  & {\sf p}[l]:= ({\sf a}, {\sf b}, {\sf d}q^l),&  (l \in \ZZ_{\bf n}) .  
\elea(ltwp)
Let ${\cal L} (s; {\sf p}', {\sf p}), \textsc{L} (t; {\sf p}', {\sf p})$ and $\textsc{L}^\dagger (t; {\sf p}', {\sf p})$ be the $L$-operators in (\req(LLp'p)), (\req(Lp'pdag)) with the parameter $(\varepsilon, \phi, \phi^\prime, \rho, \nu)$ defined in (\req(par=)). For $l, l' \in \ZZ_{\bf n}$, let $(\varepsilon^\diamond, \phi^\diamond, \phi^{' \diamond}, \rho^\diamond, \nu^\diamond)$, $(\varepsilon^\circ, \phi^\circ, \phi^{' \circ}, \rho^\circ, \nu^\circ)$ be the parameters in (\req(par=)) corresponding to $({\sf p}'(-l'), {\sf p}(l))$ , $({\sf p}'[l'], {\sf p}[l])$ respectively, related to $(\varepsilon, \phi, \phi^\prime, \rho, \nu)$  by 
\bea(lllll)
q^{\varepsilon^\diamond} = q^{\varepsilon-l}, & q^{\phi^\diamond}= q^{\phi+\frac{l-l'}{2}}, &q^{-\phi^{' \diamond}}= q^{-\phi^\prime+\frac{l-l'}{2}}, &\rho^\diamond = q^{l-l'}\rho , & \nu^\diamond =  q^{l'-l} \nu ; \\
q^{\varepsilon^\circ} = q^{\varepsilon},& q^{\phi^\circ}= q^{\phi+\frac{l'+l}{2}}, &q^{-\phi^{' \circ}}= q^{-\phi^\prime+\frac{l'+l}{2}}, & \rho^\circ = \rho , & \nu^\circ =   \nu . \\  
\elea(ppch)
Then

(i) The $\CZ^{\bf n}$-representations, $s_{\varepsilon, \phi, \phi^\prime}$ and $s_{\varepsilon^\diamond, \phi^\diamond, \phi^{' \diamond}}$ of $U_q(sl_2)$ in (\req(crep)) (or the induced ${\Large\textsl{U}}_\omega (sl_2)$-representation in (\req(crepU))), are equivalent if and only if $l-l'= 2m ~ (m \in \ZZ_{\bf n})$, 
where $s_{\varepsilon^\diamond, \phi^\diamond, 
\phi^{' \diamond}}= (\widehat{X}^{\prime m} \widehat{Z}^{\prime  -l} ) s_{\varepsilon, \phi, \phi^\prime} (\widehat{Z}^{\prime  l} \widehat{X}^{\prime  -m})$. In particular, the $L$-operators in (\req(LLp'p)) are equivalent if and only $l = l' \in \ZZ_{\bf n}$, where ${\cal L}(s; {\sf p}'(-l), {\sf p}(l)) = \widehat{Z}^{\prime  -l}{\cal L}(s; {\sf p}', {\sf p})\widehat{Z}^{\prime  l}$ , and $\textsc{L} (t; {\sf p}'(-l), {\sf p}(l)) = \widehat{Z}^{\prime  -l} \textsc{L} (t; {\sf p}', {\sf p})\widehat{Z}^{\prime  l}$.  
Furthermore, the $\CZ^N$-representations of ${\Large\textsl{U}}_\omega (sl_2)$ in (\req(crXZ)) for $(\varepsilon, \phi, \phi^\prime)$ and $(\varepsilon^\diamond, \phi^\diamond, \phi^{' \diamond})$ are equivalent if and only if $l=2n, l-l'=2m \in 2\ZZ_{\bf n}$, where $(\req(crXZ))_{\varepsilon^\diamond, \phi^\diamond, \phi^{' \diamond}} = (\widehat{X}^m \widehat{Z}^n )(\req(crXZ))_{\varepsilon, \phi, \phi^\prime}(\widehat{Z}^{-n} \widehat{X}^{ -m})$. In particular, the $L$-operators in (\req(LtauC)) are equivalent if and only if $l=l'$ with $\textsl{L} (t; {\sf p}'(-l), {\sf p}(l)) = \widehat{Z}^n \textsl{L} (t; {\sf p}', {\sf p})\widehat{Z}^{-n}$. 

(ii) The representations (\req(cycUdag)) of ${\Large\textsl{U}}_\omega (sl_2)$ for $(\varepsilon^\circ, \phi^\circ, \phi^{' \circ}, \rho^\circ, \nu^\circ)$ in (\req(ppch)) are all equivalent for $l', l \in \ZZ_{\bf n}$; the same for the $L$-operators $\textsc{L}^\dagger (t; {\sf p}'[l'], {\sf p}[l])$ in (\req(Lp'pdag)) with the relation $\textsc{L}^\dagger (t; {\sf p}'[l'], {\sf p}[l])= \widehat{X}^{' -(l'+l)} \textsc{L}^\dagger (t; {\sf p}', {\sf p})\widehat{X}^{' (l'+l)}$.  Furthermore, the equivalence of $\CZ^N$-representations (\req(crXZ)) for $(\varepsilon^\circ, \phi^\circ, \phi^{' \circ})$ and $L$-operators $\textsl{L} (t; {\sf p}'[l'], {\sf p}[l])$ in (\req(LtauC)) is given by the condition: $l'+l = 2m \in 2\ZZ_{\bf n}$, where  $\textsl{L} (t; {\sf p}'[l'], {\sf p}[l])= \widehat{X}^m \textsl{L} (t; {\sf p}', {\sf p})\widehat{X}^{-m} $.
\end{lem}
$\Box$ \par \vspace{.1in} \noindent
{\bf Remark.} (I) In Lemma \ref{lem:l-p} $(i)$, the representations $s_{\varepsilon^\diamond, \phi^\diamond, \phi^{' \diamond}}$  for $({\sf p}'(-l'), {\sf p}(l))$ and $({\sf p}'(-l'+N), {\sf p}(l+N))$ are equivalent with the same $\rho^\diamond , ~ \nu^\diamond $. The parameter $({\sf p}'_-, {\sf p}_-)$ in (\req(Lp'p-)) is equal to $({\sf p}'(-1), {\sf p}(1))$ with ${\cal L}(s), \textsc{L} (t)$-relation in (\req(LXXZg)). When ${\bf n}= N$ odd, $s_{\varepsilon^\diamond, \phi^\diamond, \phi^{' \diamond}}$ for $l, l' \in \ZZ_{\bf n}$ are all equivalent  with $m = (l-l') (\frac{{\bf n}+1}{2}) \in \ZZ_{\bf n}$. However, in ${\bf n}= 2N$ case, the requirement of the constraint $l-l'= 2m$ for equivalent representations in (\req(crep)) is necessary. Furthermore, in ${\bf n}=N$ odd case, one has $\textsl{L} (t; {\sf p}'(-l), {\sf p}(l)) = \widehat{Z}^\frac{l(N+1)}{2} \textsl{L} (t; {\sf p}', {\sf p})\widehat{Z}^\frac{-l(N+1)}{2}$, hence $\textsl{L} (t; {\sf p}'_\pm , {\sf p}_\pm)$ are equivalent. When ${\bf n}=2N$, since $1 \not\in 2\ZZ_{\bf n}$, $\textsl{L}({\sf p}_\pm', {\sf p}_\pm; t)$ are not equivalent with non-isomorphic ${\Large\textsl{U}}_\omega (sl_2)$-representation (\req(crXZ)).
\par \noindent
(II) The $({\sf p}^{\prime \dagger}_- , {\sf p}^\dagger_-)$ in (\req(pp'dag)) is equal to $({\sf p}'[0], {\sf p}[1])$ in Lemma \ref{lem:l-p} $(ii)$. Hence $\textsl{L} (t; {\sf p}^{\prime \dagger}_-, {\sf p}^\dagger_-)= \widehat{X}^\frac{N+1}{2} \textsl{L} (t; {\sf p}^{\prime \dagger}_+, {\sf p}^\dagger_+)\widehat{X}^{\frac{-(N+1)}{2}}$ for ${\bf n}=N$ odd case. When ${\bf n}=2N$, $\textsl{L} (t; {\sf p}^{\prime \dagger}_\pm, {\sf p}^\dagger_\pm)$ are not equivalent.
\par \vspace{.1in} \noindent

For convenience, we shall also use the following convention to identify the index $\pm$ in Lemma \ref{lem:Cpm} with $\ZZ_2$:
\bea(lll)
\gamma = \pm := \pm 1  = (-1)^i & \leftrightarrow & i = \frac{1-\gamma}{2} =0, 1 \in \ZZ_2. 
\elea(pm01)
For a positive integer $L$, $\{ \pm \}^L$ will also be identified with $\ZZ_2^L$ via
$$
\vec{i} = (i_1, \ldots, i_L) \in \ZZ_2^L ~ ~ \leftrightarrow ~ ((-1)^{i_1}, \ldots, (-1)^{i_L}) \in \{\pm \}^L ,
$$
in particular, $\vec{0} = (0, \ldots, 0) \leftrightarrow (+, \ldots, +), \vec{1} = (1, \ldots, 1) \leftrightarrow (-, \ldots, -)$. We shall also denote $\vec{i+1}:= \vec{i} + \vec{1} \in \ZZ_2^L$.

\subsection{The $\tau^{(2)}$-model and XXZ-model with cyclic representation  \label{ssec.XXZc}}
For a chain of size $L$, we consider the $\tau^{(2)}$-model (\req(Tranf)) defined by $L$-operator (\req(LtauC)) with  
parameters $\{ ({\sf p'}_{i_\ell} , {\sf p}_{i_\ell}) \}_{\ell=1}^L$: 
\bea(ll)
\bigotimes_{\ell=1}^L   L(t ;{\sf p}_{i_\ell}' , {\sf p}_{i_\ell})  =  \left( \begin{array}{cc}
        A (t)  & B(t) \\
        C(t) & D(t)
\end{array} \right), & 
 \tau^{(2)}(t ; \{ {\sf p'}_{i_\ell} \}, \{ {\sf p}_{i_\ell} \}) = A (t) + \omega^r D(t),
\elea(t2pp's)
satisfying the boundary condition 
\bea(lll)
| \sigma_{L+1}\rangle = | \sigma_1 - r \rangle    ~ ~ & \bigg(\Leftrightarrow &  \widehat{| k_{L+1}} \rangle   = \omega^{-r k_1} \widehat{| k_1} \rangle   \bigg),
\elea(tBy)
and the periodic-parameter condition $({\sf p'}_{L+1}, {\sf p}_{L+1})= ({\sf p'}_1, {\sf p}_1)$.  
Then $\tau^{(2)}(t ; \{ {\sf p'}_{i_\ell} \}, \{ {\sf p}_{i_\ell} \})$ commutes with the charge operator $\widehat{Z} (:= \prod_{\ell} \widehat{Z}_\ell)$, which is the same as the spin-shift operator $X (:= \prod_{\ell} X_\ell)$ with the eigenvalue $\omega^Q$ for $Q \in \ZZ_N$. The XXZ-model and $\tau^{(2)}$-model in (\req(Tranf)) with the monodromy matrix (\req(MXXZ)), (\req(Mtau)) and $L$-operator (\req(LLp'p)) will be denoted by 
\bea(lll)  
\stackrel{L}{\bigotimes} {\cal L}(s; {\sf p}'{\sf p} ) =  \left( \begin{array}{cc}
        {\cal A} (s)  & {\cal B} (s) \\
        {\cal C} (s) & {\cal D} (s)
\end{array} \right), &
 {\cal T}(s) (= {\cal T}(s; {\sf p'}, {\sf p})) = {\cal A} (s) + q^{-r'} {\cal D} (s) ; \\
\stackrel{L}{\bigotimes}  \textsc{L} (t; {\sf p}', {\sf p} )  =  \left( \begin{array}{cc}
        {\textsc{A}} (t)  & {\textsc{B}} (t) \\
        {\textsc{C}} (t) & {\textsc{D}} (t)
\end{array} \right), &
\textsc{t}^{(2)}(t)(= \textsc{t}^{(2)}(t; {\sf p'}, {\sf p}))= {\textsc{A}} (\omega t)+ q^{-r'} {\textsc{D}} (\omega t), 
\elea(XXZcyc)
satisfying the boundary condition 
\bea(lll)
| \sigma_{L+1}\rangle \rangle = | \sigma_1 + r' \rangle \rangle   ~ ~ & \bigg(\Leftrightarrow &  \widehat{| k_{L+1}} \rangle \rangle  = q^{r' k_1} \widehat{| k_1} \rangle \rangle   \bigg).
\elea(XXZBy)
Then $[{\cal T}(s), \widehat{Z}']=[\textsc{t}^{(2)}(t), \widehat{Z}^{' -2}] = 0$, where $\widehat{Z}' = \prod_\ell \widehat{Z}'_\ell ~ (= X' = \prod_\ell X'_\ell)$. The eigenvalue of $\widehat{Z}'$ and  $ \widehat{Z}^{' -2}$ will be denoted by $q^{Q'}, \omega^Q $ for $ Q' \in \ZZ_{\bf n}, Q \in \ZZ_N$  respectively. By (\req(tauT)) and (\req(LLrel)), the transfer matrices in (\req(XXZcyc)) are related by
\bea(lll)
\textsc{t}^{(2)}(t) = (-sq^{-1})^L (\frac{\omega {\sf a'a c^2}}{\sf  b^{' 3}b^3})^\frac{L}{4}\widehat{Z}^{'-1}  {\cal T} (q^{-1}s)   & \Leftrightarrow &
{\cal T} (s)  = (-s)^L (\frac{\omega {\sf a'a c^2}}{\sf  b^{' 3}b^3})^\frac{-L}{4}\widehat{Z}' \textsc{t}^{(2)}(\omega^{-1} t). 
\elea(Lrelcy)
Similarly, we denote the $\tau^{(2)}$-model in (\req(Tranf)) with the $L$-operator  (\req(Lp'pdag)) and the boundary condition  (\req(XXZBy)) by 
\bea(lll)  
\stackrel{L}{\bigotimes} \textsc{L}^\dagger (t; {\sf p}', {\sf p} )= \left( \begin{array}{cc}
        {\textsc{A}}^\dagger (t)  & {\textsc{B}}^\dagger (t) \\
        {\textsc{C}}^\dagger (t) & {\textsc{D}}^\dagger (t)
\end{array} \right), &
\textsc{t}^{\dagger (2)}(t) (= \textsc{t}^{\dagger (2)}(t; {\sf p'}, {\sf p})) ={\textsc{A}}^\dagger(\omega t)  + \omega^{-r'}{\textsc{D}}^\dagger(\omega t). 
\elea(t2dag)
The $ \widehat{Z}^{' -2}$-eigenvalue of $\textsc{t}^{\dagger (2)}(t)$ will be denoted by $\omega^Q $.

In this subsection, we study the relation between the $\tau^{(2)}$- and XXZ-model in (\req(t2pp's)), (\req(XXZcyc)) and (\req(t2dag)). For $\textsc{t}^{(2)}(t), {\cal T}(s) $ in (\req(XXZcyc)), we assume the boundary condition in (\req(XXZBy)) is related to (\req(tBy)) by $r'\equiv 2r {\pmod{\bf n}}$. First, we derive the relation between (\req(t2pp's)) and XXZ-model in (\req(XXZcyc)). For $\vec{i} = (i_1, \ldots, i_L) \in \ZZ_2^L$ with $\ZZ_2$ identified with $\pm$ in (\req(pm01)), we define the following sub-quantum space of $\textsc{t}^{(2)}$-model:
\bea(ll)
{\cal C}^{ \vec{i}}= \otimes_\ell {\cal C}_\ell \subseteq \bigotimes^L \CZ^{\bf n} , & ({\cal C}_\ell := {\cal C}^{(-1)^{i_\ell}}), \\
 \wp_{\vec{i}} (:=\otimes_\ell \wp_{(-1)^{i_\ell}}): {\cal C}^{ \vec{i}} \simeq \bigotimes^L \CZ^N ,& |\widehat{k}_1, \ldots , \widehat{k}_L \rangle \rangle_{\vec{i}} (:= \otimes_\ell \widehat{|k_\ell} \rangle \rangle_{(-1)^{i_\ell}})   \mapsto |\widehat{k}_1, \ldots, \widehat{k}_L \rangle ,
\elea(Cis)
where ${\cal C}^\pm, \wp_\pm $ are in (\req(Cpm)) (\req(wp)). Then ${\cal C}^{ \vec{i}}$ is a sub-representation of the ABCD algebra of $\textsc{t}^{(2)}$-model in (\req(XXZcyc)). By Lemma \ref{lem:Cpm}, $\wp_{\vec{i}}$ in (\req(Cis)) induces an equivalence between the $\textsc{t}^{(2)}$-monodromy matrix on ${\cal C}^{\vec{i}}$ and  ${\tau}^{(2)}$-monodromy matrix (\req(t2pp's)) with 
\be
({\sf p}_{i_\ell}', {\sf p}_{i_\ell}) := ({\sf p'}_{(-1)^{i_\ell}}, {\sf p}_{(-1)^{i_\ell}} ),  
\ele(pXXZ)
where $({\sf p}_\pm', {\sf p}_\pm )$ are defined in (\req(Lp'p)) (\req(Lp'p-)). Note that the parameter in (\req(crXZ))  for the  $\CZ^N$-representations of ${\Large\textsl{U}}_\omega (sl_2)$  associated to $\textsl{L}({\sf p}_\pm', {\sf p}_\pm; t)$  differ only in $\varepsilon$ and $\varepsilon-1$, in particular with the same $\nu$ in (\req(Mtau)). 
Through $\wp_{\vec{i}}$ in (\req(Cis)), $\textsc{t}^{(2)}(t)$ on ${\cal C}^{\vec{i}}$  is equivalent to $\tau^{(2)}(t ; \{ {\sf p'}_{i_\ell} \}, \{ {\sf p}_{i_\ell} \})$ via
\bea(l)
 \textsc{t}^{(2)}(t)_{|{\cal C}^{\vec{i}}} =  \wp_{\vec{i}}^{-1}\cdot \tau^{(2)}(t ; \{ {\sf p'}_{i_\ell} \}, \{ {\sf p}_{i_\ell} \}) \cdot \wp_{\vec{i}}, 
\elea(t2p'p)
with the relation $\widehat{Z}^{' -2} = \wp_{\vec{i}}^{-1}\cdot \widehat{Z} \cdot \wp_{\vec{i}}$. 
Indeed by Lemma \ref{lem:Cpm}, the structure of $\textsc{t}^{(2)}$-model is given by
\begin{prop}\label{prop:T2cyl} 
Let $\textsc{t}^{(2)}(t)$ be the $\tau^{(2)}$-model in (\req(XXZcyc)) with the boundary condition $r'=2r$ in (\req(XXZBy)), and $({\sf p}_{i_\ell}', {\sf p}_{i_\ell})$ the parameter defined in (\req(pXXZ)). Then 

(i) When ${\bf n}=N$ odd, $\bigotimes^L \CZ^{\bf n} = {\cal C}^{\vec{i}}$ for $\vec{i} \in \ZZ_2^L$, as representations of ABCD algebra in (\req(Mtau)), and $\textsc{t}^{(2)}(t) \simeq \tau^{(2)}(t ; \{ {\sf p'}_{i_\ell} \}, \{ {\sf p}_{i_\ell} \})$ via (\req(t2p'p)). The equivalent relations of ${\cal C}^{\vec{i}}$'s are induced by the isomorphism $
{\cal C}^{\vec{0}} \simeq {\cal C}^{\vec{i}}$: $|\widehat{k}_1, \ldots , \widehat{k}_L \rangle \rangle_{\vec{0}} \mapsto |\widehat{k}_1, \ldots , \widehat{k}_L \rangle \rangle_{\vec{i}}$.

(ii) When ${\bf n}=2N$, $\bigotimes^L \CZ^{\bf n} = \bigoplus_{\vec{i} \in\ZZ_2^L} {\cal C}^{\vec{i}}$ as representations of ABCD algebra in (\req(Mtau)), hence relations in (\req(t2p'p)) give rise to the isomorphism $\textsc{t}^{(2)}(t) \simeq \bigoplus_{\vec{i} \in\ZZ_2^L}  \tau^{(2)}(t ; \{ {\sf p'}_{i_\ell} \}, \{ {\sf p}_{i_\ell} \})$.
\end{prop}
$\Box$ \par \vspace{.1in} \noindent
We now use results obtained in Proposition \ref{prop:T2cyl} to study the XXZ-model in (\req(XXZcyc)). By (\req(Lrelcy)), the transfer matrix ${\cal T} (s)$ differs from $\textsc{t}^{(2)}(t)$ by a scale factor and a multiple of $\widehat{Z}'$, which interchanges ${\cal C}^{\vec{i}}$ and ${\cal C}^{\vec{i+1}}$ by (\req(Cpm=)). Indeed, $\widehat{Z}'$ identifies the $\textsc{t}^{(2)}$-transfer matrix on these subspaces as follows:
\begin{lem}\label{lem:t2Cj+1} 
$\textsc{t}^{(2)}(t)_{|{\cal C}^{\vec{i}}} = \widehat{Z}^{' -1} \textsc{t}^{(2)}(t)_{|{\cal C}^{\vec{i+1}}}  \widehat{Z}'$. As a consequence, the $\textsc{t}^{(2)}$-eigenvectors $v_{\vec{i}} \in  {\cal C}^{\vec{i}}$ and $ v_{\vec{i+1}} \in  {\cal C}^{\vec{i+1}}$ with the same eigenvalue are related by $v_{\vec{i+1}}= \widehat{Z}'( v_{\vec{i}})$ (up to a non-zero scale). 
\end{lem}
{\it Proof.} By (\req(XZpm)), $\widehat{Z}'$ induces an one-to-one correspondence between ${\cal C}^{\vec{i}}$ and ${\cal C}^{\vec{i+1}}$, which is related to the $(\otimes^L \CZ^N)$-automorphism  $\prod_\ell \widehat{Z}^{-i_\ell}_\ell$ via the projection $\wp_{\vec{i}}$'s in (\req(Cis)) as follows:
$$
\begin{array}{clcclc}
 {\cal C}^{\vec{i}} &\stackrel{\widehat{Z}'}{\longrightarrow}& {\cal C}^{\vec{i+1}} , &  |\widehat{k}_1, \ldots , \widehat{k}_L \rangle \rangle_{\vec{i}} &\mapsto & \omega^{- \sum_\ell i_\ell k_\ell } |\widehat{k}_1, \ldots , \widehat{k}_L \rangle \rangle_{\vec{i+1}} \\
\wp_{\vec{i}} \downarrow & &  \downarrow \wp_{\vec{i+1}}&  \downarrow & &  \downarrow  \\
\otimes^L \CZ^N & \stackrel{\prod_\ell \widehat{Z}^{-i_\ell}_\ell}{\longrightarrow}& \otimes^L \CZ^N ,&|\widehat{k}_1, \ldots , \widehat{k}_L \rangle  &\mapsto & \omega^{- \sum_\ell i_\ell k_\ell } |\widehat{k}_1, \ldots , \widehat{k}_L \rangle . 
\end{array}
$$
By (\req(Lgaequ)), one obtains the identification
$$
\tau^{(2)}(t ; \{ {\sf p'}_{i_\ell} \}, \{ {\sf p}_{i_\ell} \}) = (\prod_\ell \widehat{Z}^{i_\ell}_\ell) \tau^{(2)}(t ; \{ {\sf p'}_{i_\ell+1} \}, \{ {\sf p}_{i_\ell+1} \}) (\prod_\ell \widehat{Z}^{-i_\ell}_\ell).
$$
Indeed, the gauge relation in (\req(Lgaequ)) corresponds to the action of $\widehat{Z}'$ of $\stackrel{L}{\bigotimes} \CZ^{\bf n}$, hence follows the result.
$\Box$ \par \vspace{.1in} \noindent

\begin{prop}\label{prop:XXZT} 
Let ${\cal T} (s)$ be the XXZ-model (\req(XXZcyc)) with the boundary condition $r'=2r$ in (\req(XXZBy)), and $({\sf p}_{i_\ell}', {\sf p}_{i_\ell})$ be the parameters in (\req(pXXZ)). Then

(i) When ${\bf n}=N$ odd, $\bigotimes^L \CZ^{\bf n} = {\cal C}^{\vec{i}}$ for $\vec{i} \in (\ZZ_2)^L$, as representations of XXZ-(ABCD-)algebra in (\req(MXXZ)). The XXZ-transfer matrix ${\cal T} (s)$ in (\req(XXZcyc)) is related to $\tau^{(2)}(\omega^{-1} t ; \{ {\sf p'}_{i_\ell} \}, \{ {\sf p}_{i_\ell} \})$ by 
$$
{\cal T} (s)  =  \wp_{\vec{i}}^{-1}\cdot (-s)^L (\frac{\omega {\sf a'a c^2}}{\sf  b^{' 3}b^3})^\frac{-L}{4}\widehat{Z}^\frac{N-1}{2} \tau^{(2)}(\omega^{-1} t ; \{ {\sf p'}_{i_\ell} \}, \{ {\sf p}_{i_\ell} \}) \cdot  \wp_{\vec{i}}.
$$ 

(ii) When ${\bf n}=2N$,  we define ${\cal C}^{[\vec{i}]} = {\cal C}^{\vec{i}} \oplus {\cal C}^{\vec{i+1}}$ for a coset $[\vec{i}] (:=\{ \vec{i}, \vec{i+1} \}) \in \ZZ_2^L/\langle \vec{1} \rangle$. Then one has the decomposition of $\bigotimes^L \CZ^{\bf n}$ as representations of XXZ-algebra (\req(MXXZ)):
$$
\bigotimes^L \CZ^{\bf n} = \bigoplus \{ {\cal C}^{[\vec{i}]} ~ | ~  [\vec{i}] \in  \ZZ_2^L/\langle \vec{1} \rangle \} .
$$
The transfer matrices ${\cal T}(s)$ and $\textsc{t}^{(2)}(t)$ on each ${\cal C}^{[\vec{i}]}$ are related by (\req(Lrelcy)).  Indeed, ${\cal T}$-eigenvectors in  ${\cal C}^{[\vec{i}]}$ are  $v_{\vec{i}} \pm q^{-Q} v_{\vec{i+1}}$ with the eigenvalue $\pm q^Q(-s)^L (\frac{\omega {\sf a'a c^2}}{\sf  b^{' 3}b^3})^\frac{-L}{4}\textsc{t}^{(2)}(\omega^{-1} t)$ and $\ZZ_{\bf n}$-charge $Q, Q+N$ respectively, where $v_{\vec{i}}, v_{\vec{i+1}}$ are the $\textsc{t}^{(2)}$-eigenvectors in Lemma \ref{lem:t2Cj+1} with the $\textsc{t}^{(2)}$-eigenvalue $\textsc{t}^{(2)}(t)$ and $\ZZ_N$-charge $Q$. 
\end{prop}
{\it Proof.} $(i)$ follows easily from Proposition \ref{prop:T2cyl} $(i)$ and (\req(Lrelcy)). When ${\bf n}=2N$, $\widehat{Z}'$ interchanges the factors in ${\cal C}^{[\vec{i}]}$. Therefore  ${\cal C}^{[\vec{i}]}$ is a component of $\bigotimes^L \CZ^{\bf n}$ as representations of XXZ-algebra. The result about ${\cal T}$-eigenvectors in ${\cal C}^{[\vec{i}]}$ follows from Lemma \ref{lem:t2Cj+1}, where $\textsc{t}^{(2)}$-eigenvectors $v_{\vec{i}}, v_{\vec{i+1}}$ are related by $\widehat{Z}'(v_{\vec{i}})= v_{\vec{i+1}}, \widehat{Z}'(v_{\vec{i+1}}) = \omega^{-Q} v_{\vec{i}}$ . 
$\Box$ \par \vspace{.1in} \noindent
{\bf Remark.} When the pair $(\vec{i}, \vec{i+1})$ is changed to $(\vec{i+1}, \vec{i})$ in Lemma \ref{lem:t2Cj+1},  $( v_{\vec{i}}, v_{\vec{i+1}})$ are replaced by $(v_{\vec{i+1}}, \omega^{-Q}v_{\vec{i}})$, by which the ${\cal T}$-eigenvectors $v_{\vec{i}} \pm q^{-Q} v_{\vec{i+1}}$ in Proposition \ref{prop:XXZT} is changed to $v_{\vec{i+1}} + q^Q v_{\vec{i}}$.
\par \vspace{.1in} \noindent

We now study the structure of $\textsc{t}^{\dagger (2)}$-model in (\req(t2dag)).
For $\vec{i} = (i_1, \ldots, i_L) \in \ZZ_2^L$, define the parameter and  the sub-quantum space  of $\textsc{t}^{\dagger (2)}$-model:
\bea(ll)
({\sf p}^{' \dagger}_{i_\ell}, {\sf p}^\dagger_{i_\ell}) := ({\sf p}^{' \dagger}_{(-1)^{i_\ell}}, {\sf p}^\dagger_{(-1)^{i_\ell}} ), &
{\cal C}^{\dagger \vec{i}}= \otimes_\ell {\cal C}^\dagger_\ell \subseteq \bigotimes^L \CZ^{\bf n} , ~ ~ ({\cal C}^\dagger_\ell := {\cal C}^{\dagger (-1)^{i_\ell}}), \\
 \wp^\dagger_{\vec{i}} (:=\otimes_\ell \wp^\dagger_{(-1)^{i_\ell}}): {\cal C}^{ \dagger \vec{i}} \simeq \bigotimes^L \CZ^N ,& |\sigma_1, \ldots , \sigma_L \rangle \rangle^\dagger_{\vec{i}} (:= \otimes_\ell |\sigma_\ell \rangle \rangle^\dagger_{(-1)^{i_\ell}})   \mapsto |\sigma_1, \ldots, \sigma_L \rangle, 
\elea(pp'dag)
where $({\sf p}^{' \dagger}_\pm, {\sf p}^\dagger_\pm )$  and ${\cal C}^{\dagger \pm}, \wp^{\dagger}_\pm$ are defined in (\req(pp'dag)), (\req(Cpmdag)), (\req(wpdag)) respectively. 
Similar to Proposition \ref{prop:T2cyl} and Lemma \ref{lem:t2Cj+1}, the relation between the $\tau^{(2)}$-models in (\req(t2pp's)), (\req(t2dag)) is given by  
\begin{prop}\label{prop:T2dag} 
Let $\textsc{t}^{\dagger (2)}(t)$ be the $\tau^{(2)}$-model in (\req(t2dag)) with the boundary conditions  $r'$ in (\req(XXZBy))  related to $r$ in (\req(tBy)) by $r' \equiv -r \pmod{N}$.
Then $\textsc{t}^{\dagger (2)}(t)_{|{\cal C}^{\dagger \vec{i}}} =  \wp^{\dagger -1}_{\vec{i}} \cdot \tau^{\dagger (2)}(t ; \{ {\sf p}^{' \dagger}_{i_\ell} \}, \{ {\sf p}^\dagger_{i_\ell} \}) \cdot \wp^\dagger_{\vec{i}}$,
by which $\widehat{X}^{' -2} = \wp_{\vec{i}}^{\dagger -1}\cdot \widehat{X} \cdot \wp^\dagger_{\vec{i}}$,  and the following results hold:

(i) When ${\bf n}=N$ odd, $\bigotimes^L \CZ^{\bf n} = {\cal C}^{\dagger \vec{i}}$ for $\vec{i} \in \ZZ_2^L$, as representations of ABCD algebra in (\req(Mtau)), and $\textsc{t}^{\dagger (2)}(t) \simeq \tau^{(2)}(t ; \{ {\sf p}^{' \dagger}_{i_\ell} \}, \{ {\sf p}^\dagger_{i_\ell} \})$, where the equivalent relations among ${\cal C}^{\dagger \vec{i}}$'s are induced by $
{\cal C}^{\dagger \vec{0}} \simeq {\cal C}^{\dagger \vec{i}}$: $ |\sigma_1, \ldots , \sigma_L \rangle \rangle^\dagger_{\vec{0}} \mapsto |\sigma_1, \ldots , \sigma_L \rangle \rangle^\dagger_{\vec{i}}$.

(ii) When ${\bf n}=2N$, $\bigotimes^L \CZ^{\bf n} = \bigoplus_{\vec{i} \in\ZZ_2^L} {\cal C}^{\dagger \vec{i}}$ as representations of ABCD algebra in (\req(Mtau)), hence $\textsc{t}^{\dagger (2)}(t) \simeq \bigoplus_{\vec{i} \in\ZZ_2^L}  \tau^{(2)}(t ; \{ {\sf p}^{' \dagger}_{i_\ell} \}, \{ {\sf p}^\dagger_{i_\ell} \})$.

(iii) $\textsc{t}^{\dagger (2)}(t)_{|{\cal C}^{\dagger  \vec{i}}} = \widehat{X}^{' -1} \textsc{t}^{\dagger  (2)}(t)_{|{\cal C}^{\dagger  \vec{i+1}}}  \widehat{X}'$, hence the $\textsc{t}^{\dagger  (2)}$-eigenvectors $v^\dagger_{\vec{i}} \in  {\cal C}^{\dagger  \vec{i}}$ and $ v^\dagger_{\vec{i+1}} \in  {\cal C}^{\dagger  \vec{i+1}}$ with the same eigenvalue are related by $v^\dagger_{\vec{i+1}}= \widehat{X}'( v^\dagger_{\vec{i}})$ (up to a non-zero scale). 
\end{prop}
$\Box$ \par \vspace{.1in} \noindent

\section{Duality of $\tau^{(2)}$-models and XXZ-models with cyclic representation \label{sec.t2dual}}
In this section, we discuss the duality of $\tau^{(2)}$-models with cyclic representation, as a generalization of the $\tau^{(2)}$-duality in CPM \cite{R09}.

\setcounter{equation}{0}
\subsection{$\tau^{(2)}$-duality in chiral Potts model \label{ssec.tau2dual}}
In this subsection, we recall the $\tau^{(2)}$-duality in chiral Potts model in \cite{R09}. 
Consider the $\tau^{(2)}$-model $\tau^{(2)}(t ; \{ {\sf p'}_{i_\ell} \}, \{ {\sf p}_{i_\ell} \})$ in (\req(t2pp's)) with the boundary condition (\req(tBy)). 
Then $\tau^{(2)}(t ; \{ {\sf p'}_{i_\ell} \}, \{ {\sf p}_{i_\ell} \})$ preserves the $Q$-subspace $V_{r, Q} $ of $\stackrel{L}{\bigotimes} \CZ^N$ with the following (Hermitian) orthonormal bases:
\bea(lll)
V_{r, Q} & = \bigoplus_{k_\ell } \CZ |\widehat{k}_1, \ldots, \widehat{k}_L  \rangle & (\sum_{\ell=1}^L k_\ell \equiv Q \pmod{N}, ~ ~ \widehat{k}_{L+1} \equiv  \omega^{-rk_1} \widehat{k}_1   )\\
&= \bigoplus_{n_\ell} \CZ |Q; n_1, \ldots n_L \rangle & (\sum_{\ell=1}^L n_\ell \equiv r \pmod{N}, ~ ~ n_{L+1} \equiv \omega^{-Q n_1}  n_1 ),
\elea(Vbasis)
where $| Q; n_1, \ldots n_L \rangle :=  N^{-1/2} \sum_{\sigma_1=0}^{N-1} \omega^{-Q \sigma_1} |\sigma_1, \ldots \sigma_L \rangle$ with $\sigma_\ell - \sigma_{\ell +1} = n_\ell$ (see,  e.g. \cite{AuP, AuP7, HH} in $r=0$ case, and \cite{R09, R10}). 
Define the dual correspondence between $({r, Q})$- and $(r^*, Q^*)$-spaces (with respective to local $\CZ^N$-basis $\{ |\sigma \rangle \}, \{ \widehat{|n }\rangle \}$)  (\cite{R09} (3.16)):
\bea(llll)
\Psi : V_{r, Q} \longrightarrow V_{r^*, Q^*}, & |Q; n_1, \ldots n_L \rangle \mapsto |\widehat{n}_1, \ldots, \widehat{n}_L  \rangle, & (\sum_{\ell=1}^L n_\ell \equiv r) & (r^*, Q^*) =(Q, r) ;
\elea(Psi)
and duality of parameters ${\sf p} \in \CZ^3$ (\cite{R09} (3.9)\footnote{For the discussion of the chiral Potts model, the duality of rapidities in \cite{R09}(3.9) differs here by a constant $\alpha = {\rm i}^\frac{1}{N}$ as described in Remark of Proposition \ref{prop:dut2} in this paper.}):
\be
{\sf p}= ({\sf a}, {\sf b}, {\sf d}) \longrightarrow  {\sf p}^* = ({\sf a}^*, {\sf b}^*, {\sf d}^*):= ( {\sf a} {\sf d}, {\sf b} {\sf d}^{-1}, {\sf d}^{-1}). 
\ele(pp*)
The following lemma is used in the study of  $\tau^{(2)}$-matrix and  $\tau^{(2)}$-duality in CPM (\cite{B93} (2.14)-(2.15), \cite{BBP} (3.48), \cite{R09}(3.3)-(3.9)):
\begin{lem}\label{lem:t2dual} 
Let 
$$
\begin{array}{ll}
\textsl{L}( t; {\sf p'}, {\sf p}) = \left( \begin{array}{cc}
        \textsl{L}_0^0(t ; {\sf p'}, {\sf p})   &   \textsl{L}_0^1 (t ; {\sf p'}, {\sf p})    \\
         \textsl{L}_1^0(t ; {\sf p'}, {\sf p})  & \textsl{L}_1^1(t ; {\sf p'}, {\sf p})
\end{array} \right) , & {\sf p}' = ({\sf a', b', d'}), {\sf p} = ({\sf a, b, d}), 
\end{array}
$$
be $L$-operator in (\req(LtauC)) with ${\sf p}' , {\sf p}$ in (\req(Lp'p)). Denote 
$$
\begin{array}{ll}
\textsl{L}_{m ~ \sigma}^{m' \sigma''}(t; {\sf p'}, {\sf p})= \langle \sigma| \textsl{L}_m^{m'}(t ; {\sf p'}, {\sf p})| \sigma'' \rangle , & 
\textsl{L}_{n ~ k^* }^{n' k^{* ''}}(t; {\sf p}^*, {\sf p'}^*)= \langle \widehat{k^* }| \textsl{L}_n^{n'}(t ; {\sf p}^*, {\sf p'}^*)\widehat{| k^{* ''}} \rangle ,
\end{array}
$$
where $\sigma, \sigma'', k^*, k^{* ''} \in \ZZ_N$, $m, m', n, n' =0,1$. Define
\bea(ll)
\textsl{E}({\sf p'})^{~ ~ ~ \sigma''}_{m ; ~ \sigma} := \omega^{-m \sigma} F_{\sf p'}(\sigma-\sigma'', m ), & \textsl{E}({\sf p})_{\sigma}^{\sigma''; ~ m'}:= \omega^{m'\sigma''}\frac{\eta_{\sigma-\sigma''}}{\eta_{m'}}F_{\sf p}(\sigma-\sigma'', m') \\
U_{{\sf p}, {\sf p}'}({}_a^d| {}_b^c):= \sum_{m=0, 1}\textsl{E}({\sf p})_a^{d ; ~ m} \textsl{E}({\sf p'})^{~ ~ ~ c}_{m; ~ b}
\elea(Upp')
for $a, b, c, d \in \ZZ_N$, where 
\bea(lllll)
\frac{\eta_1}{\eta_0} = - \omega t, &
F_{\sf p}(0, 0)=1, & F_{\sf p}(0, 1)= \frac{- \omega t}{\sf b}, &
F_{\sf p}(1, 0)=  \frac{\sf d }{\sf b}, &F_{\sf p}(1, 1)=  \frac{ -\omega {\sf a d }}{\sf b}, \\
 &F_{\sf p}(\alpha, m)= 0 & {\rm if} ~  \alpha \neq  0 , 1.  
\elea(F01)
Then

(i) $\textsl{L}_{m ~ \sigma}^{m'  \sigma''} (\omega t; {\sf p'}, {\sf p}) = \textsl{E}({\sf p'})^{~ ~ ~ \sigma''}_{m ;~ \sigma}\textsl{E}({\sf p})_{\sigma}^{\sigma''; ~ m'}$ for $\sigma, \sigma'' \in \ZZ_N$, $m, m'=0,1$.

(ii) $U_{{\sf p}, {\sf p}'}({}_a^d| {}_b^c)=0$ if  $a-d$ or $b-c \neq 0,1$, and 
$$
U_{{\sf p}, {\sf p}'}({}_a^d| {}_b^c) = \textsl{L}_{n ~ k^*}^{n' k^{* ''}}(\omega t; {\sf p}^*, {\sf p'}^*) ~ ~ {\rm when} ~ (a-d, b-c) = (n, n'), n, n' =0, 1, 
$$
where $a-b=k^*, d-c= k^{* ''} \in \ZZ_N$.
\end{lem}
{\it Proof.} $(i)$ follows from (\req(LtauC)) and the definition of $\textsl{E}({\sf p'})^{~ ~ ~ \sigma''}_{m; ~ \sigma}$ and $ \textsl{E}({\sf p})_{\sigma}^{\sigma''; ~ m'}$. One also finds 
$$
U_{{\sf p}, {\sf p}'}({}_a^d| {}_b^c)= \sum_{m=0, 1} \omega^{m(d-b)} (-\omega t)^{a-d-m} F_{\sf p}(a-d , m) F_{\sf p'}(b-c, m)
$$
whose non-zero values are determined by entries of 
$$
\textsl{L} (\omega t ; {\sf p}^*, {\sf p'}^*) = \left( \begin{array}{cc}
        1  -  \omega t \frac{1 }{{\sf b} {\sf b'}} \widehat{Z}   & (\frac{{\sf d'}}{{\sf b'}}  -\omega   \frac{{\sf a'}{\sf d'} }{{\sf b} {\sf b'}} \widehat{Z}) \widehat{X}^{-1} \\
       - \omega t  (\frac{\sf d}{{\sf b} } -  \frac{{\sf a}{\sf d}}{{\sf b} {\sf b'}}\widehat{Z})\widehat{X} & - \omega t \frac{{\sf d} {\sf d'}}{ {\sf b} {\sf b'}}+  \omega \frac{{\sf a} {\sf a'} {\sf d} {\sf d'}}{{\sf b} {\sf b'}}\widehat{Z} 
\end{array} \right).
$$
Then $(ii)$ follows.
$\Box$ \par \vspace{.1in} \noindent
Using Lemma \ref{lem:t2dual} and (\req(Upp')), one finds the product form of $\tau^{(2)}(t; \{ {\sf p'}_{i_\ell} \}, \{ {\sf p}_{i_\ell} \})$ (\cite{BBP} $(3.44a)_{k=0, j=2}$, \cite{B93} (2.16)):
\be
\tau^{(2)}(t ; \{ {\sf p'}_{i_\ell} \}, \{ {\sf p}_{i_\ell} \})_{\{\sigma_\ell \}}^{\{ \sigma_\ell'' \}} =  \prod_{\ell =1}^L U_{{\sf p}_\ell, {\sf p'}_{\ell+1}}({}_{\sigma_\ell}^{\sigma''_\ell} | {}_{\sigma_{\ell+1}}^{\sigma''_{\ell+1}} ), 
\ele(taupd)
where $U_{{\sf p}, {\sf p}'}({}_a^d| {}_b^c)$ are defined in (\req(Upp')). 
Note that by the boundary condition (\req(tBy)), $\textsl{E}({\sf p'}_{L+1})^{~ ~ ~\sigma''_{L+1}}_{m ; ~ \sigma_{L+1}} = \omega^{m r}  \textsl{E}({\sf p}'_1)^{~ ~ ~ \sigma''_1}_{m; ~ \sigma_1}$, which  contributes the $\omega^r$-factor in (\req(t2pp's)).
\begin{prop}\label{prop:dut2} ${\rm (\cite{R09}  ~ Proposition ~ 3.1)}$
Let ${\sf p}_\ell^*, {\sf p'}_\ell^*$ be the dual of ${\sf p}_\ell, {\sf p'}_\ell$ in (\req(pp*)), and $\Psi$ be the dual correspondence between  $V_{r, Q}$ and $V_{r^*, Q^*}$ with $(r^*, Q^*) =(Q, r)$ in (\req(Psi)). Then  
$\tau^{(2)}(t ; \{ {\sf p'}_{i_\ell} \}, \{ {\sf p}_{i_\ell} \})$ on $V_{r, Q}$ is equivalent to  $\tau^{(2)}(t ; \{ {\sf p}^*_{i_\ell} \}, \{ {\sf p'}^*_{i_\ell+1} \})$ on $V_{r^*, Q^*}$ by 
\be
\tau^{(2)}(t ; \{ {\sf p'}_{i_\ell} \}, \{ {\sf p}_{i_\ell} \})  = \Psi^{-1} \tau^{(2)}(t ; \{ {\sf p}^*_{i_\ell} \}, \{ {\sf p'}^*_{i_\ell+1} \})  \Psi .
\ele(t2d)
\end{prop}
{\it Proof.} By (\req(taupd)) and Lemma \ref{lem:t2dual}, one finds
$$
\langle Q; n_1, \ldots n_L | \tau^{(2)}(t ; \{ {\sf p'}_{i_\ell} \}, \{ {\sf p}_{i_\ell} \}) |Q; n'_1, \ldots n'_L \rangle = \langle \widehat{n}_1, \ldots \widehat{n}_L | \tau^{(2)}(t ; \{ {\sf p}^*_{i_\ell} \}, \{ {\sf p'}^*_{i_\ell+1} \})|\widehat{n}'_1, \ldots, \widehat{n}'_L  \rangle .
$$
Then the result follows. 
$\Box$ \par \vspace{.1in} \noindent
{\bf Remark.} One may modify the duality of parameters in (\req(pp*)) by defining ${\sf p}^* = ( \alpha {\sf a} {\sf d}, \alpha {\sf b} {\sf d}^{-1}, {\sf d}^{-1})$, where $\alpha$ is a scale constant. Then $U_{{\sf p}, {\sf p}'}({}_a^d| {}_b^c)_{{\rm in}~ (\req(Upp'))} = {\rm dia}[1, \alpha^{-1}] \textsl{L} (\omega t^* ; {\sf p}^*, {\sf p'}^*){\rm dia}[1, \alpha] $ with $t^* = \alpha^2 t$, by which the duality (\req(t2d)) again holds if $\tau^{(2)}(t ; \{ {\sf p}^*_{i_\ell} \}, \{ {\sf p'}^*_{i_\ell+1} \})$ in (\req(t2d)) is changed to $\tau^{(2)}(t^* ; \{ {\sf p}^*_{i_\ell} \}, \{ {\sf p'}^*_{i_\ell+1} \})$.
\par \vspace{.1in} \noindent

\subsection{Duality of $\textsc{t}^{(2)}$-models with cyclic representation \label{ssec.t2dual}}
We now study the duality of $\tau^{(2)}$-models $\textsc{t}^{(2)}(t; {\sf p'}, {\sf p}), \textsc{t}^{\dagger (2)}(t; {\sf p'}, {\sf p})$ in (\req(XXZcyc)), (\req(t2dag)) with the boundary condition (\req(XXZBy)).
 As in (\req(Vbasis)), these $\tau^{(2)}$-models
 preserve the $Q'$-subspace $V^\prime_{r', Q'}$ of $\stackrel{L}{\bigotimes} \CZ^{\bf n}$, which is generated by the following orthonormal bases:
\bea(lll)
V^\prime_{r', Q'} & = \bigoplus_{k_\ell } \CZ |\widehat{k}_1, \ldots, \widehat{k}_L  \rangle \rangle & (\sum_{\ell=1}^L k_\ell \equiv Q' {\pmod{\bf n}} , ~  \widehat{k}_{L+1} \equiv  q^{r' k_1} \widehat{k}_1   )\\
&= \bigoplus_{n_\ell} \CZ |Q'; n_1,\ldots n_L \rangle\rangle &(\sum_{\ell=1}^L n_\ell \equiv -r'{\pmod{\bf n}}, n_{L+1} \equiv q^{-Q' n_1}  n_1 ),
\elea(V'bas)
where $|Q'; n_1, \ldots n_L \rangle\rangle:=  {\bf n}^{-1/2} \sum_{\sigma_1=0}^{{\bf n}-1} q^{-Q' \sigma_1} |\sigma_1, \ldots \sigma_L \rangle \rangle$ with $\sigma_\ell - \sigma_{\ell +1} = n_\ell$ . There is 
the dual correspondence between $({r', Q'})$- and $(r^{' *}, Q^{' *})$-spaces (with respective to the local $\CZ^{\bf n}$-basis $\{ |\sigma \rangle\rangle \}, \{ \widehat{|n }\rangle \rangle  \}$)  with $(r^{'*}, Q^{'*}) =(-Q, -r)$:
\bea(lll)
\Psi^\prime : V^\prime_{r', Q'} \longrightarrow V^\prime_{r^{'*}, Q^{'*}}, & |Q'; n_1, \ldots n_L \rangle \rangle \mapsto |\widehat{n}_1, \ldots, \widehat{n}_L  \rangle \rangle, & (\sum_{\ell=1}^L n_\ell \equiv -r {\pmod{\bf n}} ) .
\elea(Psi')
As the $\tau^{(2)}$-model in Lemma \ref{lem:t2dual}, the $L$-operator of $\textsc{t}^{(2)}$- and $ \textsc{t}^{\dagger (2)}$-model can be decomposed into a product form:
\begin{lem}\label{lem:tdagt2} 
Let 
$$
\begin{array}{ll}
{\textsc{L}}( t; {\sf p'}, {\sf p}) = \left( \begin{array}{cc}
        {\textsc{L}}^0_0(t ; {\sf p'}, {\sf p})   &   {\textsc{L}}^1_0 (t ; {\sf p'}, {\sf p})    \\
        {\textsc{L}}^0_1(t ; {\sf p'}, {\sf p})  & {\textsc{L}}^1_1(t ; {\sf p'}, {\sf p}),
\end{array} \right), & {\textsc{L}}^\dagger ( t; {\sf p'}, {\sf p}) = \left( \begin{array}{cc}
        {\textsc{L}}^{\dagger  0}_{~ 0}(t ; {\sf p'}, {\sf p})   &   {\textsc{L}}^{\dagger  1}_{~ 0} (t ; {\sf p'}, {\sf p})    \\
        {\textsc{L}}^{\dagger  0}_{~ 1}(t ; {\sf p'}, {\sf p})  & {\textsc{L}}^{\dagger  1}_{~ 1}(t ; {\sf p'}, {\sf p})
\end{array} \right) ,
\end{array}
$$
be $L$-operator in (\req(LXXZg)), (\req(Lp'pdag)) with ${\sf p}' , {\sf p}$ in (\req(Lp'p)). 
Denote 
$$
\begin{array}{ll}
{\textsc{L}}_{m ~ \sigma}^{m' \sigma''}(t; {\sf p'}, {\sf p})= \langle \langle \sigma| {\textsc{L}}_m^{m'}(t ; {\sf p'}, {\sf p})| \sigma'' \rangle \rangle  , & 
{\textsc{L}}_{~ m ~ \sigma }^{\dagger m' \sigma'' }(t; {\sf p'}, {\sf p})= \langle \langle \sigma | {\textsc{L}}_m^{\dagger m'}(t ; {\sf p'}, {\sf p}) | \sigma'' \rangle \rangle ,
\end{array}
$$
where $\sigma, \sigma''  \in \ZZ_{\bf n}$, $m, m'=0,1$. 
Define
\bea(ll)
{\textsc{E}}({\sf p'})^{~ ~ ~ \sigma''}_{m ; ~ \sigma} :=  q^{-m \sigma} {\cal F}_{\sf p'}(\frac{-\sigma+ \sigma''}{2}, m ), & {\textsc{E}}({\sf p})_{\sigma}^{\sigma''; ~ m'}:=  q^{m'\sigma''}\frac{\eta_{(-\sigma+\sigma'')/2}}{\eta_{m'}}{\cal F}_{\sf p}(\frac{-\sigma+\sigma''}{2}, m') ; \\
{\textsc{E}}^\dagger({\sf p'})^{~ ~ ~ \sigma''}_{m ; ~ \sigma} := \omega^{-m \sigma} F_{\sf p'}(\sigma-\sigma'', m ), & {\textsc{E}}^\dagger({\sf p})_{\sigma}^{\sigma''; ~ m'}:= \omega^{m'\sigma''}\frac{\eta_{\sigma-\sigma''}}{\eta_{m'}}F_{\sf p}(\sigma-\sigma'', m'); \\
{\textsc{U}}_{{\sf p}, {\sf p}'}({}_a^d| {}_b^c):= \sum_{m=0, 1}{\textsc{E}}({\sf p})_a^{d ; ~ m} {\textsc{E}}({\sf p'})^{~ ~ ~ c}_{m; ~ b}, &{\textsc{U}}^\dagger_{{\sf p}, {\sf p}'}({}_a^d| {}_b^c):= \sum_{m=0, 1}{\textsc{E}}^\dagger({\sf p})_a^{d ; ~ m} {\textsc{E}}^\dagger({\sf p'})^{~ ~ ~ c}_{m; ~ b}
\elea(UUdag)
for $a, b, c, d \in \ZZ_{\bf n}$ with $\frac{\eta_\sigma }{\eta_m} $ and $F_{\sf p}(\alpha, m)$ in (\req(F01)). Then

(i)  The $(m, m')$th entry of ${\textsc{L}}, {\textsc{L}}^\dagger$-operator for $m, m'=0,1$ are expressed by 
$$
\begin{array}{ll}
{\textsc{L}}_{m ~ \sigma}^{m'  \sigma''} (\omega t ; {\sf p'}, {\sf p}) = {\textsc{E}}({\sf p'})^{ ~ ~ ~ \sigma''}_{m ;~ \sigma}{\textsc{E}}({\sf p})_{\sigma}^{ \sigma''; ~ m'}, & {\textsc{L}}_{~ m ~ \sigma}^{\dagger m'  \sigma''} (\omega t ; {\sf p'}, {\sf p}) = {\textsc{E}}^\dagger ({\sf p'})^{ ~ ~ ~ \sigma''}_{m ;~ \sigma}{\textsc{E}}^\dagger({\sf p})_{\sigma}^{ \sigma''; ~ m'}, 
\end{array}
$$ 
where $\sigma, \sigma'' \in \ZZ_{\bf n}$.

(ii) ${\textsc{U}}_{{\sf p}, {\sf p}'}$  and ${\textsc{U}}^\dagger_{{\sf p}, {\sf p}'}$ are expressed by
\bea(l)
{\textsc{U}}_{{\sf p}, {\sf p}'}({}_a^d| {}_b^c) =\left\{ \begin{array}{ll}
\langle \langle \widehat{ k^*} |{ \textsc{L}}_{~ n}^{^\dagger n'}(\omega t; {\sf p}^*, {\sf p}^{' *})| \widehat{k^{* ''}} \rangle \rangle & {\rm if} ~ (a-d, b-c) = (-2n, -2n'), n, n' =0, 1 ; \\
0 & {\rm otherwise} .
\end{array} \right. \\
{\textsc{U}}^\dagger_{{\sf p}, {\sf p}'}({}_a^d| {}_b^c) =\left\{ \begin{array}{ll}
\langle \langle \widehat{ k^*} |{ \textsc{L}}_n^{n'}(\omega t; {\sf p}^*, {\sf p}^{' *})| \widehat{k^{* ''}} \rangle \rangle & {\rm if} ~ (a-d, b-c) = (n, n'), n, n' =0, 1 ; \\
0 & {\rm otherwise} .
\end{array} \right.
\elea(UUdag*)
where $a-b=k^*, d-c= k^{* ''} \in \ZZ_{\bf n}$, and ${\sf p}^*, {\sf p}^{' *}$ are the dual of ${\sf p}, {\sf p}'$ in (\req(pp*)). 
\end{lem}
{\it Proof.} By the definition of  ${\textsc{E}}({\sf p'}), {\textsc{E}}({\sf p})$ and ${\textsc{E}}^\dagger({\sf p'}), {\textsc{E}}^\dagger({\sf p})$ in (\req(UUdag)), the spin-operator expressions of (\req(LLp'p)) (\req(Lp'pdag)) yield  relations in $(i)$. 
$(ii)$ follows from the expression of ${\textsc{U}}_{{\sf p}, {\sf p}'}, {\textsc{U}}^\dagger_{{\sf p}, {\sf p}'}$ in (\req(UUdag)):
$$
\begin{array}{l}
{\textsc{U}}_{{\sf p}, {\sf p}'}({}_a^d| {}_b^c)
=\sum_{m=0, 1}q^{m(d-b)}(-\omega t)^{\frac{-a+d}{2}-m } {\cal F}_{\sf p}(\frac{-a+d}{2}, m) {\cal F}_{\sf p'}(\frac{-b+ c}{2}, m ), \\
{\textsc{U}}^\dagger_{{\sf p}, {\sf p}'}({}_a^d| {}_b^c) = \sum_{m=0, 1} \omega^{m(d-b)} (-\omega t)^{a-d-m} F_{\sf p}(a-d , m) F_{\sf p'}(b-c, m)
\end{array}
$$
whose non-zero values are entries of the following respective $L$-operator:
$$
\begin{array}{l}
\left( \begin{array}{cc}
        1  -  \omega t \frac{1 }{{\sf b} {\sf b'}} \widehat{Z}'   & (\frac{{\sf d'}}{{\sf b'}}  -\omega   \frac{{\sf a'}{\sf d'} }{{\sf b} {\sf b'}} \widehat{Z}') \widehat{X}^{' 2} \\
       - \omega t  (\frac{\sf d}{{\sf b} } -  \frac{{\sf a}{\sf d}}{{\sf b} {\sf b'}}\widehat{Z}')\widehat{X}^{' -2} & - \omega t \frac{{\sf d} {\sf d'}}{ {\sf b} {\sf b'}}+  \omega \frac{{\sf a} {\sf a'} {\sf d} {\sf d'}}{{\sf b} {\sf b'}}\widehat{Z}' 
\end{array} \right). = \textsc{L}^\dagger (\omega t ; {\sf p}^*, {\sf p'}^*) ; 
\\
\left( \begin{array}{cc}
        1  -  \omega t \frac{1 }{{\sf b} {\sf b'}} \widehat{Z}^{'-2}  & (\frac{{\sf d'}}{{\sf b'}}  -\omega   \frac{{\sf a'}{\sf d'} }{{\sf b} {\sf b'}}  \widehat{Z}^{' -2}) \widehat{X}^{' -1}\\
       - \omega t  (\frac{\sf d}{{\sf b} } -  \frac{{\sf a}{\sf d}}{{\sf b} {\sf b'}}\widehat{Z}^{' -2})\widehat{X}' & - \omega t \frac{{\sf d} {\sf d'}}{ {\sf b} {\sf b'}}+  \omega \frac{{\sf a} {\sf a'} {\sf d} {\sf d'}}{{\sf b} {\sf b'}}\widehat{Z}^{'-2}
\end{array} \right)= {\textsc{L}} (\omega t ; {\sf p}^*, {\sf p'}^*).
\end{array}
$$
$\Box$ \par \vspace{.1in} \noindent
Using Lemma \ref{lem:tdagt2} $(i)$ and (\req(UUdag)), one finds the product form of $\textsc{t}^{(2)}(t; {\sf p'}, {\sf p}), \textsc{t}^{\dagger (2)}(t; {\sf p'}, {\sf p})$:
$$
\begin{array}{ll}
\textsc{t}^{(2)}(t; {\sf p'}, {\sf p})_{\{\sigma_\ell \}}^{\{ \sigma_\ell'' \}} =  \prod_{\ell =1}^L {\textsc{U}}_{{\sf p}, {\sf p}'}({}_{\sigma_\ell}^{\sigma''_\ell} | {}_{\sigma_{\ell+1}}^{\sigma''_{\ell+1}} ), & 
\textsc{t}^{\dagger (2)}(t; {\sf p'}, {\sf p})_{\{\sigma_\ell \}}^{\{ \sigma_\ell'' \}} =  \prod_{\ell =1}^L {\textsc{U}}^\dagger_{{\sf p}, {\sf p}'}({}_{\sigma_\ell}^{\sigma''_\ell} | {}_{\sigma_{\ell+1}}^{\sigma''_{\ell+1}} ).
\end{array}
$$
By the boundary condition (\req(XXZBy)), one finds ${\textsc{E}}({\sf p'})^{~ ~ ~\sigma''_{L+1}}_{m ; ~ \sigma_{L+1}} = q^{-m r'}  {\textsc{E}}({\sf p'})^{~ ~ ~ \sigma''_1}_{m; ~ \sigma_1}$, ${\textsc{E}}^\dagger({\sf p'})^{~ ~ ~\sigma''_{L+1}}_{m ; ~ \sigma_{L+1}} = \omega^{-m r'}{\textsc{E}}^\dagger({\sf p'})^{~ ~ ~ \sigma''_1}_{m; ~ \sigma_1} $,  which  provide the factor of the second term in (\req(XXZcyc)) or (\req(t2dag)).
By (\req(UUdag*)), one finds
$$
\begin{array}{ll}
\langle \langle Q'; n_1, \ldots n_L | \textsc{t}^{(2)}(t; {\sf p'}, {\sf p}) |Q'; n'_1, \ldots n'_L \rangle \rangle = \langle  \langle \widehat{n}_1, \ldots \widehat{n}_L | \textsc{t}^{\dagger (2)}(t; {\sf p}^*, {\sf p}^{' *})|\widehat{n}'_1, \ldots, \widehat{n}'_L  \rangle \rangle ; \\
\langle \langle Q'; n_1, \ldots n_L | \textsc{t}^{\dagger (2)}(t; {\sf p'}, {\sf p}) |Q'; n'_1, \ldots n'_L \rangle \rangle = \langle  \langle \widehat{n}_1, \ldots \widehat{n}_L | \textsc{t}^{(2)}(t; {\sf p}^*, {\sf p}^{' *})|\widehat{n}'_1, \ldots, \widehat{n}'_L  \rangle \rangle . \\
\end{array}
$$
Hence we obtain the duality between $\tau^{(2)}$-models in (\req(XXZcyc)), (\req(t2dag)) as $\tau^{(2)}(t ; \{ {\sf p'}_{i_\ell} \}, \{ {\sf p}_{i_\ell} \})$ in Proposition \ref{prop:dut2}:
\begin{prop}\label{prop:dut2dag} 
Let ${\sf p}^*, {\sf p'}^*$ be the dual of ${\sf p}, {\sf p'}$ in (\req(pp*)), and $\Psi^\prime$ be the dual correspondence between  $V^\prime_{r', Q'}$ and $V^\prime_{r^{'*}, Q^{'*}}$ with $(r^{'*}, Q^{'*}) =(-Q, -r)$ in (\req(Psi')). Then  $\textsc{t}^{(2)}(t; {\sf p'}, {\sf p}), \textsc{t}^{\dagger (2)}(t; {\sf p'}, {\sf p})$ on $V_{r', Q'}$ is equivalent to  $\textsc{t}^{\dagger (2)}(t; {\sf p}^*, {\sf p}^{' *}), \textsc{t}^{(2)}(t; {\sf p}^*, {\sf p}^{' *})$ on $V_{r^{'*}, Q^{'*}}$ respectively by 
\bea(ll)
\textsc{t}^{(2)}(t; {\sf p'}, {\sf p}) = \Psi^{\prime -1} \textsc{t}^{\dagger (2)}(t; {\sf p}^*, {\sf p}^{' *})  \Psi^\prime , & \textsc{t}^{\dagger (2)}(t; {\sf p'}, {\sf p}) = \Psi^{\prime -1} \textsc{t}^{(2)}(t; {\sf p}^*, {\sf p}^{' *})  \Psi^\prime.
\elea(tdagd)
\end{prop}
$\Box$ \par \vspace{.1in} \noindent

\subsection{Comparison of $\tau^{(2)}$-duality and $\textsc{t}^{(2)}$-duality  \label{ssec.Relt2}}
By Proposition \ref{prop:T2cyl} and \ref{prop:T2dag}, $\textsc{t}^{(2)}(t)$ and $\textsc{t}^{\dagger (2)}(t)$ can be decomposed as a sum of $\tau^{(2)}$-models via the subspaces ${\cal C}^{\vec{i}}, {\cal C}^{\dagger \vec{i}}$ of $\stackrel{L}{\bigotimes} \CZ^{\bf n}$ for $\vec{i} \in \ZZ_2^L$ in (\req(Cis)) (\req(pp'dag)) respectively. Indeed, by (\req(t2p'p)), $\textsc{t}^{(2)}(t)_{|{\cal C}^{\vec{i}}} \simeq \tau^{(2)}(t ; \{ {\sf p'}_{i_\ell-1} \}, \{ {\sf p}_{i_\ell} \})$ with $({\sf p'}_{i_\ell}, {\sf p}_{i_\ell})$ in (\req(pXXZ)) and the boundary condition  $r' \equiv 2r {\pmod{\bf n}}$. Similarly, $\textsc{t}^{\dagger (2)}(t)_{|{\cal C}^{\dagger \vec{i}}} \simeq \tau^{\dagger (2)}(t ; \{ {\sf p}^{' \dagger}_{i_\ell} \}, \{ {\sf p}^\dagger_{i_\ell} \})$ with $({\sf p}^{' \dagger}_{i_\ell}, {\sf p}^\dagger_{i_\ell})$ in (\req(pp'dag)) and the boundary condition $r' \equiv -r \pmod{N}$. One can lift the basis of $V_{r, Q} $ in (\req(Vbasis)) to ${\cal C}^{\vec{i}}_{r, Q}$ and ${\cal C}^{\dagger \vec{i}}_{r, Q}$, denoted by 
\bea(lll)
{\cal C}^{\vec{i}}_{r, Q} & = \bigoplus_{k_\ell } \CZ |\widehat{k}_1, \ldots, \widehat{k}_L  \rangle \rangle_{\vec{i}} & = \bigoplus_{n_\ell} \CZ |Q; n_1, \ldots n_L \rangle \rangle_{\vec{i}} ,  \\
{\cal C}^{\dagger \vec{i}}_{r, Q} & = \bigoplus_{k_\ell } \CZ |\widehat{k}_1, \ldots, \widehat{k}_L  \rangle \rangle^\dagger _{\vec{i}} & = \bigoplus_{n_\ell} \CZ |Q; n_1, \ldots n_L \rangle \rangle^\dagger_{\vec{i}} ,  \\
\elea(CvVrQ)
with $\sum_{\ell=1}^L k_\ell \equiv Q , \sum_{\ell=1}^L n_\ell \equiv r \pmod{N}$, where $|\widehat{k}_1, \ldots, \widehat{k}_L  \rangle \rangle_{\vec{i}}$, $|Q; n_1, \ldots n_L \rangle \rangle_{\vec{i}}$, $|\widehat{k}_1, \ldots \widehat{k}_L  \rangle \rangle^\dagger_{\vec{i}}$, $|Q; n_1, \ldots n_L \rangle \rangle^\dagger_{\vec{i}}$ are basis elements corresponding to $|\widehat{k}_1, \ldots, \widehat{k}_L \rangle, |Q; n_1, \ldots n_L \rangle$ of $V_{r, Q}$ via the projection $\wp_{\vec{i}}, \wp^\dagger_{\vec{i}}$ in (\req(Cis)) (\req(pp'dag)). By the expression of ${\cal C}^\pm, {\cal C}^{\dagger \pm}$-basis in (\req(Cpm)) and (\req(Cpmdag)), one finds
 the relationship between the subspaces in (\req(CvVrQ)) and (\req(V'bas)) as follows:
\bea(ll)
{\cal C}^{\vec{i}}_{r, Q} \subseteq V^\prime_{r' Q'}, & {\rm only ~ if} ~ Q=Q', 2r= r' ; \\
{\cal C}^{\dagger \vec{i}}_{r, Q} \subseteq V^\prime_{r' Q'}, & {\rm only ~ if} ~  i_0=0:   r'= -r, -2Q+\sum_\ell i_\ell \equiv Q',
\elea(CrQV)
in particular, ${\cal C}^{\dagger \vec{1}}_{r, Q}$ is not contained in any $V^\prime_{r' Q'}$. Hence in ${\bf n}=2N$ case, the $\textsc{t}^{(2)}$-duality in Proposition \ref{prop:dut2dag}, with the $\tau^{(2)}$-decomposition of $\textsc{t}^{(2)}(t), \textsc{t}^{\dagger (2)}(t)$ in Proposition \ref{prop:T2cyl}, \ref{prop:T2dag}, is different from the $\tau^{(2)}$-duality in Proposition \ref{prop:dut2}. Indeed, the difference can also be seen in the inconsistency of parameter of $\tau^{(2)}$-duality in Proposition \ref{prop:dut2dag}:
$$
\begin{array}{ll}
\textsc{t}^{(2)}(t; {\sf p'}, {\sf p})_{|{\cal C}^{ \vec{1}}} \simeq \tau^{(2)}(t ; \{ {\sf p}'_-\}, \{ {\sf p}_- \}), & 
\textsc{t}^{\dagger (2)}(t; {\sf p}^*, {\sf p}^{' *})_{|{\cal C}^{\dagger \vec{1}}} \simeq  \tau^{\dagger (2)}(t ; \{ {\sf p}^{* \dagger}_- \}, \{ {\sf p'}^{* \dagger}_- \}) ; \\
\textsc{t}^{\dagger (2)}(t; {\sf p'}, {\sf p})_{|{\cal C}^{\dagger \vec{1}}} \simeq \tau^{\dagger (2)}(t ; \{ {\sf p}^{' \dagger}_-\}, \{ {\sf p}^\dagger_- \}), & \textsc{t}^{(2)}(t; {\sf p}^*, {\sf p}^{' *})_{|{\cal C}^{ \vec{1}}}   \simeq \tau^{(2)}(t ; \{ {\sf p}^*_- \}, \{ {\sf p'}^*_- \}).
\end{array}
$$ 
In ${\bf n}=N$ odd case, one can identify $\stackrel{L}{\bigotimes} \CZ^{\bf n}$ with ${\cal C}^{\vec{0}}$ or ${\cal C}^{\dagger \vec{0}}$ in Proposition \ref{prop:T2cyl} or \ref{prop:T2dag}.
\begin{prop}\label{prop:duNodd} When ${\bf n}=N$ odd, one has the identical quantum subspaces, $V^\prime_{r' Q'}= {\cal C}^{\vec{0}}_{r, Q} = {\cal C}^{\dagger \vec{0}}_{r^\dagger, Q^\dagger}$ with $Q' \equiv Q \equiv -2Q^\dagger,  r' \equiv 2r \equiv -r^\dagger \pmod{N}$, and the identification of basis elements in (\req(V'bas)), (\req(CvVrQ)):
\bea(ll)
|\widehat{k}_1, \ldots , \widehat{k}_L \rangle \rangle_{\vec{0}} = |\widehat{k}_1, \ldots , \widehat{k}_L \rangle \rangle, & |Q; n_1, \ldots n_L \rangle \rangle_{\vec{0}}= |Q'; -2n_1, \ldots -2n_L \rangle \rangle , \\
|\widehat{k}^\dagger_1, \ldots , \widehat{k}^\dagger_L \rangle \rangle^\dagger _{\vec{0}} = 
|\widehat{-2k_1}^\dagger, \ldots , \widehat{-2k_L}^\dagger \rangle \rangle, & |Q^\dagger; n_1^\dagger, \ldots n_L^\dagger \rangle \rangle^\dagger_{\vec{0}}= |Q'; n_1^\dagger, \ldots n_L^\dagger \rangle \rangle ,
\elea(Nodbas)
with $\sum_{\ell=1}^L k_\ell \equiv Q, \sum_{\ell=1}^L k^\dagger_\ell \equiv Q^\dagger$, $\sum_{\ell=1}^L n_\ell \equiv r, 
\sum_{\ell=1}^L n^\dagger_\ell \equiv r^\dagger \pmod{N}$. With the identification of $\tau^{(2)}$-models in  Proposition \ref{prop:T2cyl} and \ref{prop:T2dag}, 
$$
\begin{array}{lll}
\bigg(\tau^{(2)}(t; \{{\sf p'}\}, \{{\sf p}\}) , \tau^{(2)}(t; \{{\sf p}^* \}, \{{\sf p}^{'*} \})\bigg) \simeq 
\bigg(\textsc{t}^{(2)}(t; {\sf p'}, {\sf p}), \textsc{t}^{\dagger (2)}(t; {\sf p}^*, {\sf p}^{' *})\bigg) \simeq \bigg(\textsc{t}^{\dagger (2)}(t; {\sf p'}, {\sf p}), \textsc{t}^{(2)}(t; {\sf p}^*, {\sf p}^{' *})\bigg), 
\end{array}
$$  
the duality relations in (\req(tdagd)), (\req(t2d)) are the same.  
\end{prop}
{\it Proof.}  By (\req(Cpm)),(\req(Cpmdag)) and (\req(CrQV)),  we obtain (\req(Nodbas)). On the dual space in (\req(tdagd)), one has $V^\prime_{r^{' *} Q^{'*}}= {\cal C}^{\vec{0}}_{r^*, Q^*} = {\cal C}^{\dagger \vec{0}}_{r^{* \dagger}, Q^{* \dagger}}$ with $Q^{'*} \equiv Q^* \equiv -2Q^{* \dagger},  r^{'*} \equiv 2r^* \equiv -r^{* \dagger}$, and the identification of basis elements:
$$
\begin{array}{ll}
|\widehat{k}^*_1, \ldots , \widehat{k}^*_L \rangle \rangle_{\vec{0}} = |\widehat{k}^*_1, \ldots , \widehat{k}^*_L \rangle \rangle, & |Q^*; n^*_1, \ldots n^*_L \rangle \rangle_{\vec{0}}= |Q^{'*}; -2n^*_1, \ldots -2n^*_L \rangle \rangle , \\
|\widehat{k}^{* \dagger}_1, \ldots , \widehat{k}^{* \dagger}_L \rangle \rangle^\dagger _{\vec{0}} = 
|\widehat{-2k_1^{* \dagger}}, \ldots , \widehat{-2k_L^{* \dagger}} \rangle \rangle, & |Q^{* \dagger}; n_1^{* \dagger}, \ldots n_L^{* \dagger} \rangle \rangle^\dagger_{\vec{0}}= |Q^{'*}; n_1^{* \dagger}, \ldots n_L^{* \dagger} \rangle \rangle ,
\end{array}
$$
with $\sum_{\ell=1}^L k^*_\ell \equiv Q^*, \sum_{\ell=1}^L k^{* \dagger}_\ell \equiv Q^{* \dagger}$, $\sum_{\ell=1}^L n^*_\ell \equiv r^*, \sum_{\ell=1}^L n^{* \dagger}_\ell \equiv r^{* \dagger}$. Then the following conditions are  equivalent,
$$
(r^{' *}, Q^{'*}) = (-Q, -r) \Leftrightarrow  (r^{* \dagger}, Q^{* \dagger}) = (Q, r ) \Leftrightarrow (r^*, Q^*) = (Q^\dagger, r^\dagger). 
$$
Then the dual correspondence $\Psi^\prime$ in (\req(Psi')) becomes 
$$
\begin{array}{ll}
|Q; n_1, \ldots n_L \rangle \rangle_{\vec{0}} \mapsto |n_1, \ldots , n_L \rangle \rangle^\dagger _{\vec{0}},  & 
|Q^\dagger; n_1^\dagger, \ldots n_L^\dagger \rangle \rangle^\dagger_{\vec{0}} \mapsto 
|n_1^\dagger, \ldots , n_L^\dagger \rangle \rangle_{\vec{0}},
\end{array}
$$
by which, the dualities (\req(tdagd)) and (\req(t2d)) are equivalent. 
$\Box$ \par \vspace{.1in} \noindent
Other than the situation in Proposition \ref{prop:duNodd}, the $\tau^{(2)}$-duality (\req(t2d)) in Proposition \ref{prop:dut2} can not be lifted to the duality between $\textsc{t}^{(2)}(t; {\sf p'}, {\sf p})$ and $\textsc{t}^{\dagger (2)}(t; {\sf p}^*, {\sf p}^{' *})$, neither between $\textsc{t}^{(2)}(t; {\sf p'}, {\sf p})$ and $\textsc{t}^{(2)}(t; {\sf p}^*, {\sf p'}^*)$. In the latter case, the $\tau^{(2)}$-model dual to (\req(t2p'p)) is $\tau^{(2)}(t ; \{ {\sf p}^*_{i_\ell} \}, \{ {\sf p'}^*_{i_{\ell+1}} \})$, whose $L$-operator at $\ell$th site in (\req(t2pp's)) is defined by the parameter  
$$
({\sf p}^*_{i_\ell}, {\sf p'}^*_{i_{\ell+1}})= ({\sf p}^*_+, {\sf p'}^*_+ ), ({\sf p}^*_-, {\sf p'}^*_- ), ({\sf p}^*_+, {\sf p'}^*_- ) ~ {\rm or} ~ ({\sf p}^*_-, {\sf p'}^*_+ ), 
$$
i.e. the $l$-twist (\req(ltwp)) of ${\sf p}^*$ and ${\sf p'}^*$ with the identification ${\sf p}^*_{(-1)^l} = {\sf p}^*(l), {\sf p'}^*_{(-1)^l} ={\sf p'}^*(-l)$ for $l=0,1$. Hence $\tau^{(2)}(t ; \{ {\sf p}^*_{i_\ell} \}, \{ {\sf p'}^*_{i_{\ell+1}} \})$ is not a component of the decomposition of $\textsc{t}^{(2)}(t; {\sf p}^*, {\sf p'}^*)$ in Proposition \ref{prop:T2cyl}. Since XXZ-model is related to $\textsc{t}^{(2)}$-model by (\req(Lrelcy)), the $\tau^{(2)}$-duality in Proposition \ref{prop:dut2} can not be lifted to a duality among XXZ-models. Indeed, we shall shown in the next subsection, the dual model of XXZ-model ${\cal T}(s; {\sf p'}, {\sf p})$ is given by $\textsc{t}^{(2)}(t; {\sf p}^*, {\sf p'}^*)$, so the essence of duality lies in the duality among $\tau^{(2)}$-models.

\subsection{Dual model of XXZ-model with cyclic representation \label{ssec.XXZdual}}
In this subsection, we derive the dual model of the XXZ-model ${\cal T}(s)$ in (\req(XXZcyc)) with the boundary condition  (\req(XXZBy)). As the $L$-operator of $\textsc{t}^{(2)}$-model in Lemma \ref{lem:tdagt2}, the $L$-operator of XXZ-model ${\cal T}(s)$ can be decomposed in the following product form:
\begin{lem}\label{lem:XXZLd} 
Let 
$$
\begin{array}{ll}
{\cal L}( s; {\sf p'}, {\sf p}) = \left( \begin{array}{cc}
        {\cal L}_+^+(s ; {\sf p'}, {\sf p})   &   {\cal L}_+^- (s ; {\sf p'}, {\sf p})    \\
         {\cal L}_-^+(s ; {\sf p'}, {\sf p})  & {\cal L}_-^-(s ; {\sf p'}, {\sf p})
\end{array} \right) , & \textsc{L}^\dagger (t; {\sf p}^*, {\sf p}^{'*})
 = \left( \begin{array}{cc}
        \textsc{L}_{~+}^{\dagger+}(t ; {\sf p}^*, {\sf p}^{'*})   &   \textsc{L}_{~+}^{\dagger-} (t ; {\sf p}^*, {\sf p}^{'*})    \\
         \textsc{L}_{~-}^{\dagger+}(t; {\sf p}^*, {\sf p}^{'*})  & \textsc{L}_{~-}^{\dagger-}(t; {\sf p}^*, {\sf p}^{'*})
\end{array} \right) 
\end{array}
$$
be the $L$-operator in (\req(LLp'p)), (\req(Lp'pdag)) respectively, where ${\sf p}' = ({\sf a', b', d'}), {\sf p} = ({\sf a, b, d})$ and ${\sf p}^*, {\sf p'}^*$ are the dual of ${\sf p}, {\sf p'}$ in (\req(pp*)), (here we use $\gamma= \pm (= \pm 1)$ in (\req(pm01)) as the auxiliary index of the $L$-operator). 
Denote 
$$
\begin{array}{ll}
{\cal L}_{\gamma ~ \sigma}^{\gamma' \sigma''}(s)= \langle \langle \sigma| {\cal L}_\gamma ^{\gamma'}(s ; {\sf p'}, {\sf p})| \sigma'' \rangle \rangle , & \textsc{L}_{~ \delta ~ k^*}^{\dagger \delta' k^{* ''}}(t)= \langle \langle \widehat{k^*}| {\cal L}_{~\delta}^{\dagger \delta'}(t ; {\sf p}^*, {\sf p}^{'*})|\widehat{ k^{* ''}} \rangle \rangle 
\end{array}
$$
where $\sigma, \sigma'', k^*, k^{* ''} \in \ZZ_{\bf n}$, $\gamma, \gamma', \delta, \delta' = \pm $. Define
\bea(ll)
{\cal E}({\sf p'})^{~ ~ ~ \sigma''}_{\gamma; ~ \sigma} &:=  ({\bf b'b})^\frac{m}{2} q^{-m \sigma} (-s)^{-(1+m)} {\cal F}_{\sf p'}(\sigma-\sigma'', \gamma ), ~ ~ ~ ~ (m= \frac{1-\gamma}{2}) ,\\
{\cal E}({\sf p})_{\sigma}^{\sigma''; ~ \gamma'}&:= ({\bf b'b})^\frac{-m'}{2} q^{m'(\sigma''+1)}(-s)^{m'} \frac{\eta_{\sigma-\sigma''}}{\eta_{\gamma'}}{\cal F}_{\sf p}(\sigma-\sigma'', \gamma'), ~ ~ (m'= \frac{1-\gamma'}{2}); \\
{\cal U}_{{\sf p}, {\sf p}'}({}_a^d| {}_b^c)&:= \sum_{\gamma = \pm }{\cal E}({\sf p})_a^{d ; ~ \gamma} {\cal E}({\sf p'})^{~ ~ ~ c}_{\gamma ; ~ b} , \\
\elea(XXZU)
for $a, b, c, d \in \ZZ_{\bf n}$, where $\frac{\eta_-}{\eta_+} = -s^2 $, ${\cal F}_{\sf p}(+, +)= (\frac{q{\bf b}^3}{{\sf a}{\sf d}^2})^\frac{1}{4}$, ${\cal F}_{\sf p}(+, -)= (\frac{-s^2}{\sf b})(\frac{q{\bf b}^3}{{\sf a}{\sf d}^2})^\frac{1}{4} $, ${\cal F}_{\sf p}(-, +)=   (\frac{\sf d }{\sf b})(\frac{q{\bf b}^3}{{\sf a}{\sf d}^2})^\frac{1}{4}$, ${\cal F}_{\sf p}(-, -)=  (\frac{ -\omega {\sf a d }}{\sf b})(\frac{q{\bf b}^3}{{\sf a}{\sf d}^2})^\frac{1}{4} $, and ${\cal F}_{\sf p}(\alpha, \gamma)= 0$ if $\alpha \neq  \pm 1$. Then

(i) ${\cal L}_{\gamma ~ \sigma}^{\gamma' \sigma''}(s) = {\cal E}({\sf p'})^{~ ~ ~ \sigma''}_{\gamma ;~ \sigma}{\cal E}({\sf p})_{\sigma}^{\sigma''; ~ \gamma'}$ for $\sigma, \sigma'' \in \ZZ_{\bf n}$, $\gamma, \gamma'= \pm $.

(ii) ${\cal U}_{{\sf p}, {\sf p}'}({}_a^d| {}_b^c)=0$ if  $a-d$ or $b-c \neq \pm 1$, and  
$$
\begin{array}{ll}
{\cal U}_{{\sf p}, {\sf p}'}({}_a^d| {}_b^c) = -s^{-1} (\frac{q^2{\bf b}^{* 3}{\bf b'}^{* 3}}{{\sf a}^*{\sf a'}^*{\sf c}^{*2}})^\frac{1}{4} \textsc{L}_{~ \delta ~ k^*}^{\dagger \delta' k^{* ''}}(t) & {\rm when} ~ (a-d, b-c) = (\delta, \delta'), \delta, \delta' =\pm 1 ,
\end{array}
$$
where $k^*=a-b, k^{* ''} =d-c \in \ZZ_{\bf n}$.   
\end{lem}
{\it Proof.} $(i)$ follows from  the spin-operator expression of ${\cal L} (s; {\sf p}', {\sf p})$ in (\req(LLp'p)) and the definition of ${\cal E}({\sf p'})^{~ ~ ~ \sigma''}_{\gamma; ~ \sigma}$ and ${\cal E}({\sf p})_{\sigma}^{\sigma''; ~ \gamma'}$. One also finds 
$$
{\cal U}_{{\sf p}, {\sf p}'}({}_a^d| {}_b^c)
= -s^{-1} \sum_{\gamma = \pm} q^\frac{(1-\gamma)(d-b+1)}{2} (-s^2)^\frac{\gamma -a+d}{2} {\cal F}_{\sf p}(a-d , \gamma ) {\cal F}_{\sf p'}(b-c, \gamma),
$$
of which the non-zero values form the following $L$-operator
$$
-s^{-1} (\frac{q^2{\bf b}^3{\bf b'}^3}{{\sf aa'}{\sf c}^2})^\frac{1}{4} \left( \begin{array}{cc}
        1  -  t \frac{1 }{{\sf b} {\sf b'}} \widehat{Z}'   & (\frac{{\sf d'}}{{\sf b'}}  -\omega   \frac{{\sf a'}{\sf d'} }{{\sf b} {\sf b'}} \widehat{Z}') \widehat{X}^{' 2} \\
      - t  (\frac{\sf d}{{\sf b} } -  \frac{{\sf a}{\sf d}}{{\sf b} {\sf b'}}\widehat{Z}')\widehat{X}^{' -2} & - t \frac{{\sf d} {\sf d'}}{ {\sf b} {\sf b'}}+  \omega \frac{{\sf a} {\sf a'} {\sf d} {\sf d'}}{{\sf b} {\sf b'}}\widehat{Z}' \end{array} \right)= -s^{-1} (\frac{q^2{\bf b}^{* 3}{\bf b'}^{* 3}}{{\sf a}^*{\sf a'}^*{\sf c}^{*2}})^\frac{1}{4} \textsc{L}^\dagger (t; {\sf p}^*, {\sf p}^{'*}).  
$$
Then $(ii)$ follows.
$\Box$ \par \vspace{.1in} \noindent
By Lemma \ref{lem:XXZLd} $(i)$, ${\cal T}(s; {\sf p'}, {\sf p})$ can be expressed in the product form:
$$
{\cal T}(s; {\sf p'}, {\sf p})_{\{\sigma_\ell \}}^{\{ \sigma_\ell'' \}} =  \prod_{\ell =1}^L {\cal U}_{{\sf p}, {\sf p}'}({}_{\sigma_\ell}^{\sigma''_\ell} | {}_{\sigma_{\ell+1}}^{\sigma''_{\ell+1}} ), 
$$
with the relation ${\cal E}({\sf p'})^{~ ~ ~\sigma''_{L+1}}_{\gamma ; ~ \sigma_{L+1}} = q^{-\frac{(1-\gamma)}{2} r'}  {\cal E}({\sf p'})^{~ ~ ~ \sigma''_1}_{\gamma; ~ \sigma_1}$ by (\req(XXZBy)), which  contributes the $q^{-r'}$-factor in (\req(XXZcyc)). Then Lemma \ref{lem:XXZLd} $(ii)$ yields the relation
$$
\begin{array}{ll}
&\langle \langle Q'; n_1, \ldots n_L | {\cal T}(s; {\sf p'}, {\sf p}) |Q'; n'_1, \ldots n'_L \rangle \rangle \\
= &(-s)^{-L} (\frac{q^2{\bf b}^{* 3}{\bf b'}^{* 3}}{{\sf a}^*{\sf a'}^*{\sf c}^{*2}})^\frac{L}{4} 
\langle  \langle \widehat{n}_1, \ldots \widehat{n}_L | \textsc{t}^{\dagger (2)}(\omega^{-1}t; {\sf p}^*, {\sf p}^{' *})|\widehat{n}'_1, \ldots, \widehat{n}'_L  \rangle \rangle . 
\end{array}
$$
Hence up to a scale function,  $\textsc{t}^{\dagger (2)}(\omega^{-1}t; {\sf p}^*, {\sf p}^{' *})$ is the dual model of  ${\cal T}(s; {\sf p'}, {\sf p})$: 
\begin{prop}\label{prop:duXXZ} 
Let ${\sf p}^*, {\sf p'}^*$ be the dual of ${\sf p}, {\sf p'}$ in (\req(pp*)), and $\Psi^\prime$ be the dual correspondence between  $V^\prime_{r', Q'}$ and $V^\prime_{r^{'*}, Q^{'*}}$ with $(r^{'*}, Q^{'*}) =(-Q, -r)$ in (\req(Psi')). Then  ${\cal T}(s; {\sf p'}, {\sf p})$ on $V_{r', Q'}$ is equivalent to  $(-s)^{-L} (\frac{q^2{\bf b}^{* 3}{\bf b'}^{* 3}}{{\sf a}^*{\sf a'}^*{\sf c}^{*2}})^\frac{L}{4} \textsc{t}^{\dagger (2)}(\omega^{-1}t; {\sf p}^*, {\sf p}^{' *})$ on $V_{r^{'*}, Q^{'*}}$  by 
\bea(ll)
{\cal T}(s; {\sf p'}, {\sf p}) = (-s)^{-L} (\frac{q^2{\bf b}^{* 3}{\bf b'}^{* 3}}{{\sf a}^*{\sf a'}^*{\sf c}^{*2}})^\frac{L}{4}  \Psi^{\prime -1} \textsc{t}^{\dagger (2)}(\omega^{-1} t; {\sf p}^*, {\sf p}^{' *})  \Psi^\prime .
\elea(XXZdu)
\end{prop}
$\Box$ \par \vspace{.1in} \noindent
By Proposition \ref{prop:dut2dag},  $\textsc{t}^{\dagger (2)}(t; {\sf p'}, {\sf p})$ is also the dual model of $\textsc{t}^{(2)}(t; {\sf p'}, {\sf p})$, which is related to ${\cal T}(s; {\sf p'}, {\sf p})$ by (\req(Lrelcy)). Note that the duality in (\req(taupd)) or (\req(tdagd)) is the identification of dual $\tau^{(2)}$-transfer matrices, not as the ABCD-algebra representation in (\req(Mtau)). In particular, the operator $\widehat{Z}'$ in (\req(Lrelcy)) is not preserved under the duality correspondence $\Psi^\prime$ in (\req(tdagd)).  This explains the reason why $\textsc{t}^{\dagger (2)}(t; {\sf p'}, {\sf p})$ serves the dual model for both $\textsc{t}^{(2)}(t; {\sf p'}, {\sf p})$ and ${\cal T}(s; {\sf p'}, {\sf p})$.

\section{Inhomogeneous Chiral Potts Model and XXZ model with cyclic $U_q(sl_2)$-representation \label{sec.CPM}}
\setcounter{equation}{0}
Let $\omega, q$ be roots of unity in (\req(qomega)). For convenience, we may assume $\omega = {\rm e}^\frac{2 \pi {\rm i}}{N}$, and $q = - \omega^\frac{-1}{2}$ when ${\bf n}=N$ odd, and $q = \pm \omega^\frac{-1}{2}$ when ${\bf n}= 2N$,  where $\omega^\frac{1}{2}:= {\rm e}^\frac{\pi {\rm i}}{N}$. By the gauge equivalence and the scaling of spectral parameter, the five parameters in the $L$-operator of $\tau^{(2)}$-model in (\req(LtauC)) (\req(Lp'p) ) can be reduced to the 3-parameter family represented by two rapidities in the same $k'$-curve of CPM model \cite{R0710}: 
\bea(lll)
({\sf a}',  {\sf b}', {\sf a}, {\sf b}, {\sf c})= ( x_{p'}, 
y_{p'} , x_p , y_p , \mu_{p'}\mu_p ), &
{\sf p}'= p' = (x_{p'}, y_{p'}, \mu_{p'} ), & {\sf p} = p=(x_p , y_p  , \mu_p ) ,  
\elea(CPpp)
with $p, p' \in {\goth W}_{k'}$, defined by 
\bea(ll)
{\goth W}_{k'}(= {\goth W}_{k', k}) &: k x^N  = 1 -  k'\mu^{-N},   k  y^N  = 1 -  k'\mu^N, \ ( k'^2 \neq 0, 1,   k^2 + k'^2 = 1 ) 
\elea(cpmC)
(see, e.g. \cite{AMP, BBP}). Here, ${\goth W}_{k'}$ is represented by ${\goth W}_{k', k}$ or ${\goth W}_{k', -k}$, which are isomorphic via the transformation $(x, y, \mu) \mapsto (\pm \omega^\frac{1}{2}x, \pm \omega^\frac{1}{2}y, \mu)$.

\subsection{Duality in chiral Potts model \label{ssec.duCPM}}
In chiral Potts model,  the $\tau^{(2)}$-model (\req(t2pp's)) are defined by the $L$-operators with the parameter ${\sf p'}_{i_\ell}= p'_{i_\ell} , {\sf p}_{i_\ell} = p{i_\ell} \in {\goth W}_{k'}$ in (\req(cpmC)):
\bea(lll)
\tau^{(2)}(t) = \tau^{(2)}(t ; \{ p'_{i_\ell} \}, \{ p_{i_\ell} \}), & t:= x_q y_q ~ & {\rm for} ~ ~ q \in {\goth W}_{k'}.
\elea(t2p)
It is known that the $Q$-operator of $\tau^{(2)}$-model (\req(t2p)) in the theory of Baxter's $TQ$-relation \cite{Bax} is the $\stackrel{L}{\otimes} \CZ^N$-transfer matrix in CPM \cite{B93, BBP, R075, R0710}\footnote{We shift the index of $p'_\ell$ in \cite{R09} (2.16) by one, where $p'_\ell$ is equal to $p'_{\ell+1}$ of (\req(ThatT)) here.} :
\bea(ll)
T (q)_{\{\sigma \}, \{\sigma'\}} &(= T (q ; \{ p'_\ell \}, \{ p_\ell  \} )_{\{\sigma \}, \{\sigma'\}}) = \prod_{\ell =1}^L W_{p_\ell q}(\sigma_\ell - \sigma'_\ell ) \overline{W}_{p'_{\ell+1}   q}(\sigma_{\ell+1} - \sigma'_\ell), \\
\widehat{T} (q)_{\{\sigma' \}, \{\sigma'' \}}&(=\widehat{T} (q ; \{ p'_\ell \}, \{ p_\ell  \} )_{\{\sigma' \}, \{\sigma''\}}) = \prod_{\ell =1}^L \overline{W}_{p_\ell q}(\sigma'_\ell - \sigma''_\ell) W_{p'_{\ell+1}  q}(\sigma'_\ell - \sigma''_{\ell+1}),
\elea(ThatT)
where  $q \in {\goth W}_{k'}$ in (\req(cpmC)), $\sigma, \sigma', \sigma''  \in \ZZ_N$ are the spin-basis in (\req(Fb)), and  $W_{p q}, \overline{W}_{p q}$ are the Boltzmann weights in CPM \cite{BPA}:
\be
\frac{W_{p q}(\sigma)}{W_{p q}(0)}  = (\frac{\mu_p}{\mu_q})^\sigma \prod_{j=1}^\sigma
\frac{y_q-\omega^j x_p}{y_p- \omega^j x_q }  , \ ~ \
\frac{\overline{W}_{p q}(\sigma)}{\overline{W}_{p q}(0)}  = ( \mu_p\mu_q)^\sigma \prod_{j=1}^\sigma \frac{\omega x_p - \omega^j x_q }{ y_q- \omega^j y_p } 
\ele(WW)
with $W_{p,q}(0)= \overline{W}_{p,q}(0)=1$. Here both (\req(t2p)) and  (\req(ThatT)) are with the boundary condition (\req(tBy)) and the periodic-vertical rapidities  $(p'_{L+1}, p_{L+1})= (p'_1, p_1)$. The star-triangle relation of Boltzmann weights \cite{AMPT, AuP, BPA,  MaS, MPTS}  yields 
$$
T (q) \widehat{T} (r) = (\frac{f_{p' q}f_{p r}}{f_{p q}f_{p' r}})^L T (r) \widehat{T} (q) , \ \ \widehat{T} (q) T (r) = (\frac{f_{p q}f_{p' r}}{f_{p' q}f_{p r}})^L  \widehat{T} (r) T (q) ,
$$
with the commutative relations, 
$$[\widehat{T} (q) T (r),  \widehat{T} (q') T (r')] = [T (q) \widehat{T} (r), T (q') \widehat{T} (r')] = 0$$ 
for $q, r, q', r' \in {\goth W}_{k'}$, where $f_{pq} =  (\frac{{\rm det}_N(\overline{W}_{p q}(i-j))}{\prod_{n=0}^{N-1} W_{pq}(n)})^{1/N}$. Then $T (q), \widehat{T}(q)$ can be diagonalized via two invertible $q$-independent matrices $P_B, P_W$, i.e. $P_W^{-1} T (q) P_B=  T_{\rm diag} (q)$, $P_B^{-1} \widehat{T}(q)P_W = \widehat{T}_{\rm diag} (q) $ diagonal so that $\widehat{T}_{\rm diag} (q) =   T_{\rm diag} (q) (\frac{f_{p q} }{f_{p' q}})^L D$ for some $q$-independent diagonal matrix $D$ (\cite{BBP} (2.32)-(2.34), (4.46), \cite{B93} (2.10)-(2.13). The Baxter's $TQ$-relation is in the form of $\tau^{(2)}T$-relation between (\req(t2p)) and (\req(ThatT)) (\cite{B93} (3.15), \cite{BBP} (4.31), \cite{R0805} (2.31)-(2.32)):  
\bea(l)
\tau^{(2)}(t_q) T (Uq) =  \varphi_q T( q) + \omega^r \overline{\varphi}_{Uq} X T (U^2 q) ,  \\
\tau^{(2)}(t_q) T(U'q) = \omega^r \varphi_q^\prime  X T(q) + \overline{\varphi}_{U'q}^\prime T(U'^2 q) , 
\elea(tauTU)
where $U, U'$ are automorphisms of (\req(cpmC)) defined by $U(x,y,\mu)=(\omega x,y,\mu)$, $ U'(x,y,\mu)=(x,\omega y,\mu)$, and $\varphi_q (= \varphi_{\{p'_\ell \}, \{p_\ell \}; q}), \overline{\varphi}_q (=\overline{\varphi}_{\{p'_\ell \}, \{p_\ell \};q})$, $\varphi_q^\prime(= \varphi_{\{p'_\ell \}, \{p_\ell \};q}^\prime ), \overline{\varphi}_q^\prime(=\overline{\varphi}_{\{p'_\ell \}, \{p_\ell \}; q}^\prime )$ are functions defined by
$$
\begin{array}{ll}
\varphi_q   = \prod_{\ell} \frac{(t_{p'_\ell}- t_q)  (y_{p_\ell}-  \omega x_q)}{y_{p_\ell} y_{p'_\ell}(x_{p'_\ell}-  x_q)}, & \overline{\varphi}_q =
\prod_{\ell} \frac{\omega \mu_{p'_\ell} \mu_{p_\ell}(t_{p_\ell}- t_q)(x_{p'_\ell}- x_q) }{y_{p_\ell} y_{p'_\ell}(y_{p_\ell}- \omega x_q)}  , \\
\varphi_q^\prime  =  
\prod_{\ell} \frac{\omega \mu_{p_\ell} \mu_{p'_\ell}(t_{p'_\ell}- t_q)(x_{p_\ell}- y_q) }{y_{p_\ell} y_{p'_\ell}(y_{p'_\ell}- y_q)}, &  
\overline{\varphi}_q^\prime  = \prod_{\ell} \frac{(t_{p_\ell}- t_q)  (y_{p'_\ell}-  y_q)}{y_{p_\ell} y_{p'_\ell}(x_{p_\ell} -  y_q)}.
\end{array}
$$

We now discuss the duality of $\tau^{(2)}$-model (\req(t2p)) in CPM. In order to preserve the vertical rapidities in (\req(t2p)), we modify the parameter-duality  (\req(pp*)) by a factor $\alpha= {\rm i}^\frac{1}{N} $ in the Remark of Proposition \ref{prop:dut2}, so that the parameter-duality defines the dual rapidities between $k'$- and $k'^{-1}$-curves ${\goth W}_{k'}$ and ${\goth W}_{1/k'}$ (\cite{R09} (3.9)):
\be
{\goth W}_{k'} \stackrel{\sim}{\longrightarrow} {\goth W}_{1/k'} , ~ p= (x_p, y_p, \mu_p)  \mapsto  p^* = (x_{p*}, y_{p*}, \mu_{p*}):= ( {\rm i}^\frac{1}{N} x_p \mu_p, {\rm i}^\frac{1}{N} y_p \mu_p^{-1}, \mu_p^{-1}). 
\ele(pp*ch)
Then the $\tau^{(2)}$-duality (\req(t2d)) takes the following form:  
\bea(ll)
\tau^{(2)}(t ; \{ p'_{i_\ell} \}, \{ p_{i_\ell} \})  = \Psi^{-1} \tau^{(2)}(t^* ; \{ p^*_{i_\ell} \}, \{ p^{' *}_{i_\ell+1} \})  \Psi , & t^* = (-1)^\frac{1}{N} t.
\elea(t2dCP)
The duality (\req(t2dCP)) can be extended to the duality of CPM. Indeed, $T, \widehat{T}$ in (\req(ThatT)) commute with the charge operator $\widehat{Z} (= X)$, hence preserve the $Q$-subspace  $V_{r, Q}$ in (\req(Vbasis)). 
The Boltzmann weights (\req(WW)) for dual rapidities in  (\req(pp*ch)) are related by $
\frac{\overline{W}^{(f)}_{p q }(k)}{\overline{W}^{(f)}_{pq}(0)} = W_{p^*q^*}(k)$ , $\frac{W^{(f)}_{p q}(k)}{W^{(f)}_{p q}(0)} =  \overline{W}_{p^*q^*}(N-k)$, where $\overline{W}^{(f)}_{pq}(k)= \frac{1}{\sqrt{N}} \sum_{\sigma = 0}^{N-1} \omega^{k \sigma }\overline{W}_{pq}(\sigma) $, $W^{(f)}_{pq}(k) = \frac{1}{\sqrt{N}} \sum_{\sigma = 0}^{N-1} \omega^{k \sigma }W_{pq}(\sigma)$ are the Fourier transform of Boltzmann weights in (\req(WW)), (\cite{R09} (3.17)). By \cite{R09} Theorem 3.1, when $(r^*, Q^*)= (Q, r)$, the chiral Potts transfer matrices (\req(ThatT)),  $T (q ; \{ p'_\ell \}, \{ p_\ell \})$, $\widehat{T} (q ; \{ p'_\ell \}, \{ p_\ell  \} ) $ on $V_{r, Q}$ and $T (q^* ; \{ p^*_\ell \}, \{ p^{'*}_{\ell+1} \})$, $\widehat{T} (q^* ; \{ p^*_\ell \}, \{ p^{' *}_{\ell+1}  \} ) $ on $V_{r^*, Q^*}$,  are equivalent via  the dual correspondence $\Psi$ in $(\req(Psi))$: 
\bea(l)
T (q^*; \{ p^*_\ell \}, \{ p^{'*}_{\ell+1} \}) =\bigg( \prod_\ell \frac{W^{(f)}_{p_\ell^{' *} ,q^*}(0)}{W^{(f)}_{p_\ell ,q}(0)} \bigg) \Psi T (q; \{ p'_\ell \}, \{ p_\ell \}) \Psi^{-1}   , \\
\widehat{T}(q^*; \{ p^*_\ell \}, \{ p^{'*}_{\ell+1} \})   = \bigg( \prod_\ell \frac{W^{(f)}_{p_\ell^* ,q^*}(0)}{W^{(f)}_{p'_\ell ,q}(0)} \bigg) \Psi \widehat{T} (q; \{ p'_\ell \}, \{ p_\ell \}) ,
\elea(TT*)
where $p^*_\ell , p^{'*}_\ell , q^* \in {\goth W}^*_{1/k'}$ are the dual of $p_\ell, p'_\ell, q \in {\goth W}_{k'}$ in (\req(pp*ch)).

\subsection{Chiral Potts transfer matrix as the $Q$-operator of XXZ-model with cyclic $U_q(sl_2)$-representation \label{ssec.CPXXZ}}
In this subsection, we study the $\textsc{t}^{(2)}$- and XXZ-model, (\req(XXZcyc)) and (\req(t2dag)) with parameter in  (\req(CPpp)): ${\sf p'}=p', {\sf p}=p \in {\goth W}_{k'}$. Then $({\sf p}_-', {\sf p}_-)$ in (\req(aminus)) and 
$({p}^{\prime \dagger}_-, {\sf p}^\dagger_-)$ in (\req(pp'dag)) become
$$
{\sf p}_-'= (q^{-1} x_{p'}, qy_{p'}, \mu_{p'}), ~ {\sf p}_- = ( q x_p , q^{-1} y_p, \mu_p) , ~
{\sf p}^{\prime \dagger}_-= p', ~ {\sf p}^\dagger_-= (x_p, y_p , q \mu_p) ,
$$
which are elements in ${\goth W}_{k'}$ in ${\bf n}=N$ odd case, denoted by $p'_-, p_-, p^{\prime \dagger}_-, p^\dagger_- $. In ${\bf n}=2N$ case, ${\sf p}_-', {\sf p}_-$ are elements in ${\goth W}_{k', -k}$, identified with $p'_-= (x_{p'}, \omega^{-1} y_{p'}, \mu_{p'}), p_-= ( \omega^{-1} x_p ,  y_p, \mu_p) \in {\goth W}_{k'} (={\goth W}_{k', k})$; and ${\sf p}^\dagger_-= p^\dagger_- \in {\goth W}_{-k'}$. Hence parameters in (\req(aminus)) and (\req(pp'dag)) are identified with 
\bea(ll)
{\bf n}=N ~ {\rm odd}:& p_-'=(q^{-1} x_{p'}, qy_{p'}, \mu_{p'}) , p_-= ( q x_p , q^{-1} y_p, \mu_p),  p^{\prime \dagger}_-= p', p^\dagger_- = (x_p, y_p , q \mu_p), \\
& (p_-', p_-, p^{\prime \dagger}_-, p^\dagger_- \in {\goth W}_{k'}) ; \\
{\bf n}= 2N  :& p_-'=(x_{p'}, \omega^{-1}  y_{p'}, \mu_{p'}) , p_-= ( \omega^{-1} x_p , y_p, \mu_p),  p^{\prime \dagger}_-= p', p^\dagger_- = (x_p, y_p , q \mu_p), \\
& (p_-', p_-, p^{\prime \dagger}_- \in {\goth W}_{k'} , p^\dagger_- \in {\goth W}_{-k'}). 
\elea(CPp-)
Over the subspace ${\cal C}^{ \vec{i}}$ of $\bigotimes^L \CZ^{\bf n}$ in (\req(Cis)) for $\vec{i}  \in \ZZ_2^L$, the equivalence of $\textsc{t}^{(2)}(t) (= \textsc{t}^{(2)}(t; p', p))$ and $\tau^{(2)}$-model in (\req(t2p'p)) becomes
\bea(ll)
 \textsc{t}^{(2)}(t)_{|{\cal C}^{\vec{i}}} \simeq  \tau^{(2)}(t ; \{ p'_{i_\ell} \}, \{ p_{i_\ell} \}), & p_{i_\ell}'= p'_{(-1)^{i_\ell}} , p_{i_\ell}= p_{(-1)^{i_\ell}} \in {\goth W}_{k'}.   
\elea(t2p'pCP)
Hence we may regard $T (q ; \{ p'_\ell \}, \{ p_\ell  \} ), \widehat{T} (q ; \{ p'_\ell \}, \{ p_\ell  \} )$ in (\req(ThatT)) as the $Q$-operator of $\textsc{t}^{(2)}(t)_{|{\cal C}^{\vec{i}}}$, with $\tau^{(2)}T$-relation induced from (\req(tauTU)). In ${\bf n}=N$  odd case, by Proposition \ref{prop:T2cyl} $(i)$, $\bigotimes^L \CZ^{\bf n} = {\cal C}^{ \vec{i}}$; then by Proposition \ref{prop:XXZT} $(i)$, the XXZ-model ${\cal T}(s, p', p)$ in (\req(XXZcyc)) can be identified with $\textsc{t}^{(2)}(t)$ up to some scale function, by which one may consider $T (q ; \{ p'_\ell \}, \{ p_\ell  \} ), \widehat{T} (q ; \{ p'_\ell \}, \{ p_\ell  \} )$ as the $Q$-operator of ${\cal T}(s, p', p)$. On the other hand, since $p^{\prime \dagger}_-, p^\dagger_- \in {\goth W}_{k'}$ by (\req(CPp-)) , the $Q$-operators of the equivalent models in Proposition \ref{prop:T2dag} $(i)$, 
$\textsc{t}^{\dagger (2)}(t)_{|{\cal C}^{\dagger \vec{i}}} \simeq \tau^{\dagger (2)}(t ; \{ p^{' \dagger}_{i_\ell} \}, \{ p^\dagger_{i_\ell} \})$ with $p^{' \dagger}_{i_\ell}=p^{' \dagger}_{(-1)^{i_\ell}}, p^\dagger_{i_\ell}= p^\dagger_{(-1)^{i_\ell}} \in {\goth W}_{k'}$, are $T (q ; \{ p^{' \dagger}_{i_\ell} \}, \{ p^\dagger_{i_\ell} \}), \widehat{T} (q ;\{ p^{' \dagger}_{i_\ell} \}, \{ p^\dagger_{i_\ell} \} )$. In particular, when $\vec{i} = \vec{0}$ where $p_{i_\ell}'= p^{' \dagger}_{i_\ell}= p' , p_{i_\ell}= p^\dagger_{i_\ell}= p$ for all $\ell$, the duality in Proposition \ref{prop:duNodd} extends to the duality of $Q$-operators with $p'_\ell = p', p_\ell =p, p^{'*}_\ell=p{'*},  p^*_\ell =p^*$in (\req(TT*)).

In ${\bf n}= 2N$  case, since $p^{\prime \dagger}_- \in {\goth W}_{k'} , p^\dagger_- \in {\goth W}_{-k'}$ in (\req(CPp-)), the vertical rapidities $p^{' \dagger}_{i_\ell},  p^\dagger_{i_\ell}$ of the $\tau^{(2)}$-model equivalent to $\textsc{t}^{\dagger (2)}(t)$ on the component ${\cal C}^{\dagger \vec{i}}$ in Proposition \ref{prop:T2dag} $(ii)$ are not all in the same $k'$-curve when $\vec{i} \neq \vec{0}$, hence the chiral Potts transfer matrix fails to be the $Q$-operator of $\textsc{t}^{\dagger (2)}$-model. For 
$\textsc{t}^{(2)}$-model in Proposition \ref{prop:T2cyl} $(ii)$, since the parameter $\nu$ in (\req(par=)) corresponding to $({\sf p'}_-, {\sf p}_-)$ in (\req(aminus)) and $(p'_-, p_-)$ in (\req(CPp-)) differ by $\omega$ in ${\bf n}= 2N$ case, $\tau^{(2)}(t ; \{ p'_{i_\ell} \}, \{ p_{i_\ell} \})$ in (\req(t2p'pCP)) and $\tau^{(2)}(t ; \{ {\sf p}'_{i_\ell} \}, \{ {\sf p}_{i_\ell} \})$  in (\req(t2p'p)) are related by the relation (\req(Ttnul)) with $\xi_\ell = \omega^{i_\ell}$.
 Therefore, the $\widehat{Z}'$-identification between $\textsc{t}^{(2)}(t)_{|{\cal C}^{\vec{i}}}$ and $\textsc{t}^{(2)}(t)_{|{\cal C}^{\vec{i+1}}}$ in Proposition \ref{prop:XXZT} $(ii)$ corresponds the identification of $\tau^{(2)}(t ; \{ p'_{i_\ell} \}, \{ p_{i_\ell} \})$ and $\tau^{(2)}(t ; \{ p'_{i_\ell+1} \}, \{ p_{i_\ell+1} \})$ in (\req(t2p'pCP)) via the change of local spectral parameters at site $\ell$, $t \mapsto \omega^{(-1)^{i_\ell+1}} t$. Then $T (q ; \{ p'_\ell \}, \{ p_\ell  \} ), \widehat{T} (q ; \{ p'_\ell \}, \{ p_\ell  \} )$ are identified  with $T (q ; \{ p'_\ell+1 \}, \{ p_\ell+1  \} ), \widehat{T} (q ; \{ p'_\ell+1 \}, \{ p_\ell+1  \} )$ with the compatible $\tau^{(2)}T$-relation (\req(tauTU)) via the change of variable $q \in {\goth W}_{k'}$ at site $\ell$: $ (x_q, y_q, \mu_q) \mapsto (x_q, \omega^{\mp 1} y_q, \mu_q)$ according to $(p'_{i_\ell} p_{i_\ell}) = (p'_\pm , p_\pm ) $. Therefore, the two pairs of chiral Potts transfer matrices,
$(T (q ; \{ p'_\ell \}, \{ p_\ell  \} ), T (q ; \{ p'_\ell+1 \}, \{ p_\ell+1  \} ))$ and $((\widehat{T} (q ; \{ p'_\ell \}, \{ p_\ell  \} ), \widehat{T} (q ; \{ p'_\ell+1 \}, \{ p_\ell+1  \} ))$, form the $Q$-operator of ${\cal T}(s; p', p)$  over the component ${\cal C}^{[i]}$ of $\stackrel{L}{\otimes} \CZ^{\bf n}$ in Proposition \ref{prop:XXZT} $(ii)$.

\section{Concluding Remarks \label{sec.F}}
In this paper, we first in section \ref{sec.t2qunt} characterize the quantum group ${\Large\textsl{U}}_\textsl{w} (sl_2)$ in the $L$-operator of $\tau^{(2)}$-model for a generic $\textsl{w}$  as a subalgebra of the quantum group $U_{\sf q} (sl_2)$ in XXZ model with ${\sf q}^{-2} = \textsl{w}$. Then we study the XXZ model with cyclic representation of $U_q (sl_2)$, and its related $\tau^{(2)}$-models in the root of unity case (\req(qomega)). Through the representation theory of ${\Large\textsl{U}}_\omega (sl_2)$, we obtain the structure of XXZ- and $\tau^{(2)}$-models, and their relationship with non-superintegrable $N$-state CPM in sections \ref{sec.CyclMod} and \ref{sec.CPM}. We also study the duality of XXZ- and $\tau^{(2)}$-models with cyclic representation in section \ref{sec.t2dual}, and find the fundamental role of $\tau^{(2)}$-models in the duality theory. One expects a similar structure will appear again in XXZ models defined by other representations of $U_q (sl_2)$, not only for the cyclic representation in this work. The representation theory should provide a more direct access to the eigenvector problem of models related to $\tau^{(2)}$-model, as indicated in \cite{R10} about the eigenvectors of superintegrable chiral Potts model. A program along this line is now under consideration, and progress is expected. We leave the discussion to future work.


\begin{thebibliography}{99}
\bibitem{AMP} G. Albertini, B. M. McCoy, and 
J. H. H. Perk, Eigenvalue spectrum of the
superintegrable chiral Potts model, In {\it Integrable system in quantum field theory and statistical mechanics,}  Adv. Stud. Pure Math., 19, Kinokuniya Academic, Academic Press, Boston, MA (1989) 1--55.
%
\bibitem{AMPT} H. Au-Yang, B. M. McCoy,  
J. H. H. Perk and S. Tang, Solvable models in statistical mechanics and Riemann surfaces of genus greater than one, {\it Algebraic Analysis}, Vol. 1 , eds. M. Kashiwara and T. Kawai, Academic Press, San Diego (1988), 29--40.
%
\bibitem{AuP} H. Au-Yang and J. H. H. Perk, Onsager's star-triangle equation: Master key to integrability, In {\it Integrable system in quantum field theory and statistical mechanics,}  Adv. Stud. Pure Math., 19, Kinokuniya Academic, Academic Press, Boston, MA (1989) 57--94.
%
\bibitem{AuP7} H. Au-Yang and J.H.H. Perk, Eigenvectors the superintegrable model I: $\goth{sl}_2$ generators, J. Phys. A: Math. Theor. 41 (2008) 275201; arXiv: 0710.5257; Eigenvectors in the superintegrable model II: ground state sector, arXiv: 0803.3029.
%
\bibitem{Bax} R. J. Baxter, Exactly solved models in statistical mechanics, Academic Press (1982).
%
\bibitem{B90} R. J. Baxter, Chiral Potts model: eigenvalues of the transfer matrix, Phys. Lett. A 146 (1990) 110--114.
%
\bibitem{B91} R. J. Baxter, Calculation of the eigenvalues of the transfer matrix of the chiral Potts model, {\it Proc. Fourth Asia-Pacific Physics Conference} (Seoul, Korea, 1990) Vol 1, World-Scientific, Singapore (1991) 42--58.
%
\bibitem{B93} R. J. Baxter, Chiral Potts model with skewed boundary conditions, J. Stat. Phys. 73 (1993) 461--495.
%
\bibitem{BBP} R. J. Baxter, V.V. Bazhanov and
J.H.H. Perk,  Functional relations for transfer
matrices of the chiral Potts model, Int. J. Mod.
Phys. B 4 (1990) 803--870.
%
\bibitem{BPA} R. J. Baxter, J. H. H. Perk and H. Au-Yang, New solutions of the  star-triangle relations for the chiral Potts model, Phys. Lett. A 128 (1988) 138--142.
%
\bibitem{BazS} V.V. Bazhanov and Yu.G. Stroganov, Chiral
Potts model as a descendant of the six-vertex model, J.
Stat. Phys. 59 (1990) 799--817.
%
\bibitem{DJMM} E. Date, M. Jimbo, K. Miki and T. Miwa, Cyclic representations of $U_q(sl(n+1, \CZ))$ at $q^N=1$, Publ. RIMS, Kyoto Univ. 27 (1991) 347--366.
%
\bibitem{DK} C. DeConcini and V. G. Kac, Representations of quantum groups at roots of unity, in {\it Operator Algebra, Unitary Representations, Enveloping Algebras, and Invariant Theory}, Paris (1989) {\it Progress in Mathematics} 92, Birkh\"{a}user, Boston, Massachusstts (1990) 471-- 506.
%
\bibitem{Fad} L. D. Faddeev, How algebraic Bethe
Ansatz works for integrable models, eds. A.
Connes, K. Gawedzki and J. Zinn-Justin, {\it Quantum symmetries/ Symmetries quantiques}, Proceedings of the Les Houches summer school, Session LXIV, Les Houches, France, August 1-September 8, 1995, North-Holland (1998),  149--219.
%
\bibitem{HH} N.S. Han and A. Honecker: Low-Temperature Expansions and Correlation Functions of the $\ZZ_3$-Chiral Potts Model, J.Phys. A27 (1994) 9-22;  BONN-HE-93-13, hep-th/9304083.
%
\bibitem{KiR} A. N. Kirillov and N. Yu. Reshetikhin, Exact solution of the integrable XXZ Heisenberg model with arbitrary spin: I. The ground state and the excitation spectrum, J. Phys. A: Math. Gen. 20 (1987) 1565 -- 1595.
%
\bibitem{KRS} P. P. Kulish,  N. Yu. Reshetikhin and E. K. Sklyanin, Yang Baxter equation and representation theory, Lett. Math. Phys. 5 (1981) 393--403.
%
\bibitem{MaS} V. B. Matveev and A. O. Simnov, Some comments on the solvable chiral Potts model, Lett. Math. Phys. 19 (1990) 179--185.
%
\bibitem{MPTS} B. M. McCoy,  
J. H. H. Perk, S. Tang and C. H. Sah, Commuting transfer matrices for the four-state self-dual chiral Potts model with a genus-three uniformizing Fermat curve, Phys. Lett. A 125 (1987) 9--14.
%
\bibitem{MR} B. M. McCoy and S. S. Roan, Excitation spectrum and phase structure of the chiral Potts model. Phys. Lett. A 150 (1990) 347--354.
%
\bibitem{R06Q} S. S. Roan, The Q-operator for root-of-unity symmetry in six vertex model, J. Phys. A: Math. Gen. 39 (2006) 12303-12325; cond-mat/0602375.
%
\bibitem{R06F} S. S. Roan, Fusion operators in the generalized $\tau^{(2)}$-model and root-of-unity symmetry of the XXZ spin chain of higher spin, J. Phys. A: Math. Theor. 40 (2007) 1481-1511; cond-mat/0607258.
%
\bibitem{R075} S. S. Roan, The transfer matrix of superintegrable chiral Potts model as the Q-operator of root-of-unity XXZ chain with cyclic representation of $U_q(sl_2)$, J. Stat. Mech. (2007) P09021; arXiv: 0705.2856.
%
\bibitem{R0710} S. S. Roan, On the equivalent theory of the generalized $\tau^{(2)}$-model and the chiral Potts model with two alternating vertical rapidities, arXiv: 0710.2764.
%
\bibitem{R0805} S. S. Roan, Bethe equation of $\tau^{(2)}$-model and eigenvalues of finite-size transfer matrix of chiral Potts model with alternating rapidities, J. Stat. Mech. (2008) P10001; arXiv:0805.1585.
%
\bibitem{R0806} S. S. Roan, On $\tau^{(2)}$-model in chiral Potts model and cyclic representation of quantum group $U_q(sl_2)$, J. Phys. A: Math. Theor. 42 (2009) 072003; arXiv:0806.0216.
%
\bibitem{R09} S. S. Roan, Duality and symmetry in chiral Potts model, J. Stat. Mech. (2009) P08012; arXiv:0905.1924.
%
\bibitem{R10} S. S. Roan, Eigenvectors of an arbitrary Onsager sector in superintegrable  $\tau^{(2)}$-model and chiral Potts model, arXiv:1003.3621.
\end{thebibliography}
\end{document}